\documentclass[a4paper,11pt]{article}
\usepackage{jcappub} 
\usepackage{lineno}
\usepackage{xcolor}
\usepackage{colortbl}
\usepackage{subcaption}

\newcommand{\jz}[1]{\textcolor{black}{#1}}
\definecolor{{C0}}{HTML}{1f77b4}

\title{\boldmath Tracing Missing Baryons in the Cosmic Filaments with tSZ and CMB-Lensing Stacking}

\author[a]{Jianzhuo Li}
\author[a,b,1]{, Yi Zheng\note{Corresponding author.}}
\author[a,b]{and Weishan Zhu}
\affiliation[a]{School of Physics and Astronomy, Sun Yat-sen University, 2 Daxue Road, Tangjia, Zhuhai, 519082, China}
\affiliation[b]{CSST Science Center for the Guangdong-Hong kong-Macau Greater Bay Area, SYSU, Zhuhai, 519082, China}

\emailAdd{lijzh86@mail2.sysu.edu.cn}
\emailAdd{zhengyi27@mail.sysu.edu.cn}
\emailAdd{zhuwshan5@mail.sysu.edu.cn}

\abstract{We investigate the distribution of missing baryons in the cosmic filaments by stacking $\sim 30,700$ filaments across the northern and southern SDSS sky regions using \textit{Planck} Compton-$y$ and CMB lensing maps. Filaments are identified using the DisPerSE algorithm applied to the SDSS LOWZ-CMASS galaxy samples, selecting structures with lengths between 30–100 cMpc and redshifts in the range $0.2 < z < 0.6$. Radial profiles are extracted out to 25 cMpc from the filament spines, and galaxy clusters with halo masses above $\sim 3 \times 10^{13}M_\odot$ are masked to reduce contamination. We detect the thermal Sunyaev–Zel’dovich (tSZ) signal at $7.82\sigma$ and the CMB lensing signal at $7.78\sigma$. The stacked profiles are corrected by a geometric bias correction based on filament inclination with respect to the line-of-sight, and they are portrayed assuming isothermal, cylindrically symmetric models. We explore different gas and matter density distributions, focusing on the $\beta$-models with $(\alpha,\beta) = (2,2/3)$ or $(1,1)$. By jointly fitting the Compton-$y$ and $\kappa$ profiles, we constrain the central electron overdensity and temperature to be $\delta = 4.18^{+2.01}_{-1.06}$ and $T_e = 2.74^{+0.65}_{-0.53}\times 10^6~\mathrm{K}$ for the standard $\beta$-model. These results suggest that filamentary WHIM in our selected long filaments contributes a significant baryon fraction of $0.127^{+0.019}_{-0.021}\times \Omega_b$ to the cosmic baryon budget.}

\begin{document}

\maketitle
\flushbottom

\section{Introduction}
\label{sec:intro}
The large-scale structure of the Universe is organized into a web-like pattern, known as the cosmic web, which comprises voids, walls, filaments, and clusters. Among these components, filaments play a crucial role by connecting galaxy clusters and are typically located at the intersections of planar structures called walls, which enclose vast, underdense regions known as voids~\cite{Klypin1983,Bond1996,deLapparent1986,Colless2001,Zehavi2011,bond2010MNRAS.409..156B,zeldovich2009A&A...500...19S,Weygaert2008,Libeskind2018MNRAS.473.1195L,Sarkar2019MNRAS.485.4743S,zhu2017ApJ...838...21Z,Darvish2017ApJ...837...16D,Hoffman2012MNRAS.425.2049H,Pfeifer2022MNRAS.514..470P,Cautun2014MNRAS.441.2923C,Arag2007A&A...474..315A,Forero-Romero2009MNRAS.396.1815F}. In dense regions such as galaxy clusters, these structures are relatively easy to detect due to the strong clustering of galaxies and the presence of hot intracluster gas. In contrast, the more diffuse filamentary components of the cosmic web are more elusive. However, with the advent of large spectroscopic redshift surveys—most notably the Sloan Digital Sky Survey (SDSS~\cite{York2000})—filamentary structures have begun to be systematically identified through the three-dimensional distribution of galaxies~\cite{Tempel2014MNRAS.438.3465T, Carr2022A&A...659A.166C, Malavasi2020A&A...642A..19M}. 

Big Bang nucleosynthesis and observations of the Cosmic Microwave Background (CMB) indicate that baryons account for approximately 5\% of the total energy density of the Universe~\cite{Cyburt2016RvMP...88a5004C,PlanckCollaboration2016}. Yet, less than half of this baryon budget has been directly observed at low redshift, residing in stars, cold gas, and the intracluster medium (ICM)~\cite{Nicastro2008Sci...319...55N,Macquart2020Natur.581..391M,Shull2012ApJ...759...23S,Fukugita1998ApJ...503..518F,Fukugita2004ApJ...616..643F,Simon2021NatAs...5..852D,Penton2004ApJS..152...29P,Paolo1992MNRAS.258P..14P}. The remaining component—commonly referred to as the “missing baryons”—is predicted to reside in the warm-hot intergalactic medium (WHIM), a diffuse, ionized gas phase with temperatures in the range of $10^5$–$10^7\,\mathrm{K}$ and overdensities of $\delta \sim 1$–$100$~\cite{Cen&Ostriker1999ApJ...514....1C,Dave2001ApJ...552..473D}, predominantly distributed along the filamentary structures of the cosmic web.

Detecting the WHIM remains challenging due to its low density and moderate temperature. Traditional probes, such as ultraviolet absorption lines and soft X-ray emission, are sensitive only to specific phases of this medium and leave a large fraction of the baryon budget unconstrained~\cite{Penton2004ApJS..152...29P,Fujita2008PASJ...60S.343F,Nicastro2008Sci...319...55N,Kull1999A&A...341...23K,Eckert2015Natur.528..105E}. The thermal Sunyaev–Zel’dovich (tSZ) effect~\cite{Zeldovich1969Ap&SS...4..301Z}, which measures the integrated electron pressure along the line-of-sight (LOS), provides a promising alternative. As it is sensitive to ionized gas across a wide range of densities and redshifts, the tSZ effect enables the detection of diffuse baryons distributed over large-scale filamentary environments.

Several studies have reported tSZ detections associated with filamentary structures. The \textit{Planck} Collaboration~\cite{PlanckCollaboration2013A&A...550A.134P} and Bonjean et al.~\cite{Bonjean2018A&A...609A..49B} detected a tSZ signal in the bridge of hot gas connecting the merging galaxy cluster pair A399–A401, indicating the presence of intercluster warm-hot gas. However, the tSZ signal from individual filaments is generally weak due to the low gas density and pressure. To enhance the signal, statistical stacking techniques have been employed. Using this approach, Tanimura et al.~(2019a)~\cite{Tanimura2019aMNRAS.483..223T} performed a stacking analysis on the regions between pairs of luminous red galaxies (LRGs) in the SDSS and detected an excess tSZ signal consistent with the presence of WHIM. Independently, de Graaff et al.~\cite{deGraff2019A&A...624A..48D} and Singari et al.~\cite{singari2020JCAP...08..028S} analyzed pairs of CMASS galaxies and reported a similar detection, further supporting the presence of diffuse warm-hot gas in filamentary environments. In a follow-up study, Tanimura et al.~(2020)~\cite{Tanimura2020A&A...637A..41T} identified filaments from the SDSS LOWZ-CMASS sample and reported a statistically significant tSZ signal in the northern Galactic sky, providing additional evidence for the presence of WHIM beyond collapsed halos.

The gravitational lensing of the CMB provides a complementary probe to understand the matter distribution in the cosmic web. The CMB lensing convergence ($\kappa$) map~\cite{PlanckCollaboration2020A&A...641A...8P} traces the projected mass distribution along the LOS. By combining tSZ and CMB lensing measurements, one can simultaneously constrain the electron pressure and total matter density, thereby breaking the degeneracy between gas density and temperature~\cite{Tanimura2020A&A...637A..41T,deGraff2019A&A...624A..48D}. This joint approach offers a unique opportunity to characterize the physical state of the WHIM in filaments.

In this work, we present a statistical detection of the WHIM in cosmic filaments using data that spans both the northern and southern regions of the SDSS sky. By jointly stacking the tSZ and CMB lensing signals across a large sample of filaments, we probe the thermodynamic and density properties of diffuse baryons distributed along the cosmic web beyond virialized halos. In turn, the key strength of our study lies in the larger sky coverage and sample size than previous studies. We stack $\sim 30,700$ filaments identified from the SDSS DR12 LOWZ-CMASS galaxy samples~\cite{Reid2016MNRAS.455.1553R} using the \textnormal{DisPerSE} algorithm~\cite{Sousbie2011MNRAS.414..350S,SousbiePichon2011MNRAS.414..384S}, with lengths between $30-100~\mathrm{cMpc}$ and redshifts in the range $0.2 < z < 0.6$. This enables improved statistical precision and allows us to test the consistency of the filament signal across different sky regions.

The structure of this paper is as follows. Section~\ref{sec:data} describes the \textit{Planck} Compton-$y$ and CMB lensing convergence maps, as well as the filament and cluster catalogs used in our analysis. Section~\ref{sec:Method} outlines the stacking methodology and validation tests. In Section~\ref{sec:model}, we model the stacked profiles using two isothermal $\beta$-models. Section~\ref{sec:results} presents the fitting results based on a Markov Chain Monte Carlo (MCMC) analysis, including constraints on the electron overdensity, core radius, temperature, and the inferred baryon budget. Section~\ref{sec:discussion} presents a detailed discussion and comparison with previous studies, and Section~\ref{sec:conclusion} concludes with a summary and outlook.

Throughout the paper, we adopt the cosmological parameters from the \textit{Planck} 2018 results~\cite{Planck2020A&A...641A...6P}, assuming a $\Lambda$CDM model with $\Omega_\mathrm{m} = 0.3153$, $\Omega_\Lambda = 0.6847$, $\Omega_\mathrm{b} = 0.0493$, and $H_0 = 67.36~\mathrm{km\,s^{-1}\,Mpc^{-1}}$.


\section{Data}
\label{sec:data}

In this section, we describe the datasets employed in our analysis, including the Compton-$y$ map from the \textit{Planck} PR4 release, the CMB lensing convergence map, and catalogs of cosmic filaments and galaxy clusters. We detail the construction of the filament catalog using the DisPerSE algorithm applied to SDSS DR12 galaxy samples, and outline the masking procedures implemented to reduce contamination from foreground emissions and massive clusters.
\subsection{Planck PR4 y map}
In this study, we utilize the latest Compton-$y$ parameter map derived from the \textit{Planck} PR4 (NPIPE) data release~\cite{Chandran2023MNRAS.526.5682C}. This full-sky $y$-map is constructed using a tailored Needlet Internal Linear Combination (NILC~\cite{Remazeilles2013MNRAS.430..370R}) pipeline applied to the nine \textit{Planck} frequency maps ranging from 30 to 857 GHz~\cite{Planck2020A&A...643A..42P}. 
Compared to the PR2 release~\cite{Planck2016A&A...594A..22P}, the PR4 NILC $y$-map demonstrates an enhanced quality: large-scale $1/f$ noise-induced striations are suppressed, Galactic thermal dust residuals are reduced, and contamination from thermal noise and the cosmic infrared background (CIB) is substantially lowered at small angular scales~\cite{Chandran2023MNRAS.526.5682C}.
\par 
The $y$-map is provided in HEALPix format~\cite{Gorski2005ApJ...622..759G}, which divides the celestial sphere into equal-area pixels. For our analysis, we adopt a resolution of $N_\mathrm{side} = 2048$ (corresponding to a pixel size of $\sim1.7$ arcmin), resulting in approximately 50 million pixels over the full sky. This high angular resolution allows for precise stacking of the tSZ signal around filamentary structures.
\par
To assess the impact of the map quality on our stacking results, we performed parallel analyses using both the PR2 and PR4 $y$-maps. The resulting stacked $y$ profiles are found to be broadly consistent between the two releases. We therefore adopt the PR4 $y$-map for all subsequent analyses presented in this work.
\par 
Prior to stacking, we apply a series of masks to the Compton-$y$ and CMB lensing maps to reduce foreground contamination. For the $y$-map, we use the NILC processing mask (NILC-MASK), which leaves $f_{\mathrm{sky}} = 0.98$; the Galactic mask from \textit{Planck} Collaboration XXII~\cite{Planck2016A&A...594A..22P} (GAL-MASK); and the point source mask (PS-MASK), retaining a combined sky fraction of $f_{\rm sky}=0.56$.

\subsection{CMB lensing convergence map}

We utilize the \textit{Planck} 2018 CMB lensing convergence ($\kappa$) map, reconstructed with a baseline minimum-variance (MV) estimator based on the spectral matching independent component analysis (SMICA) CMB map, as provided in \textit{Planck} Collaboration VIII (2020)~\cite{PlanckCollaboration2020A&A...641A...8P}. The reconstruction is performed up to a maximum multipole of $\ell_{\mathrm{max}} = 4096$. Following the \textit{Planck} team’s recommendation~\cite{PlanckCollaboration2020A&A...641A...8P} and being consistent with Tanimura et al.~(2020)~\cite{Tanimura2020A&A...637A..41T}, we restrict our analysis to the multipole range $8 \leq \ell \leq 400$, which has been demonstrated to be robust against systematic uncertainties in the \textit{Planck} 2018 release. To further reduce noise contributions from small angular scales ($\ell > 400$), we apply an exponential filter to suppress the high-multipole modes.

To reduce contamination from foreground sources, we apply the \textit{Planck}-provided masks, which exclude point sources, the Galactic plane, and regions around galaxy clusters identified in the 2015 SZ catalog~\cite{Planck2015resultsXXVI.2016A&A...594A..26P}. After masking, approximately 67\% of the sky remains available for analysis.

\subsection{Filament catalog}
\label{sec:fil_catalog}
\begin{figure}[t!]
    \centering
    \includegraphics[width=0.6\textwidth]{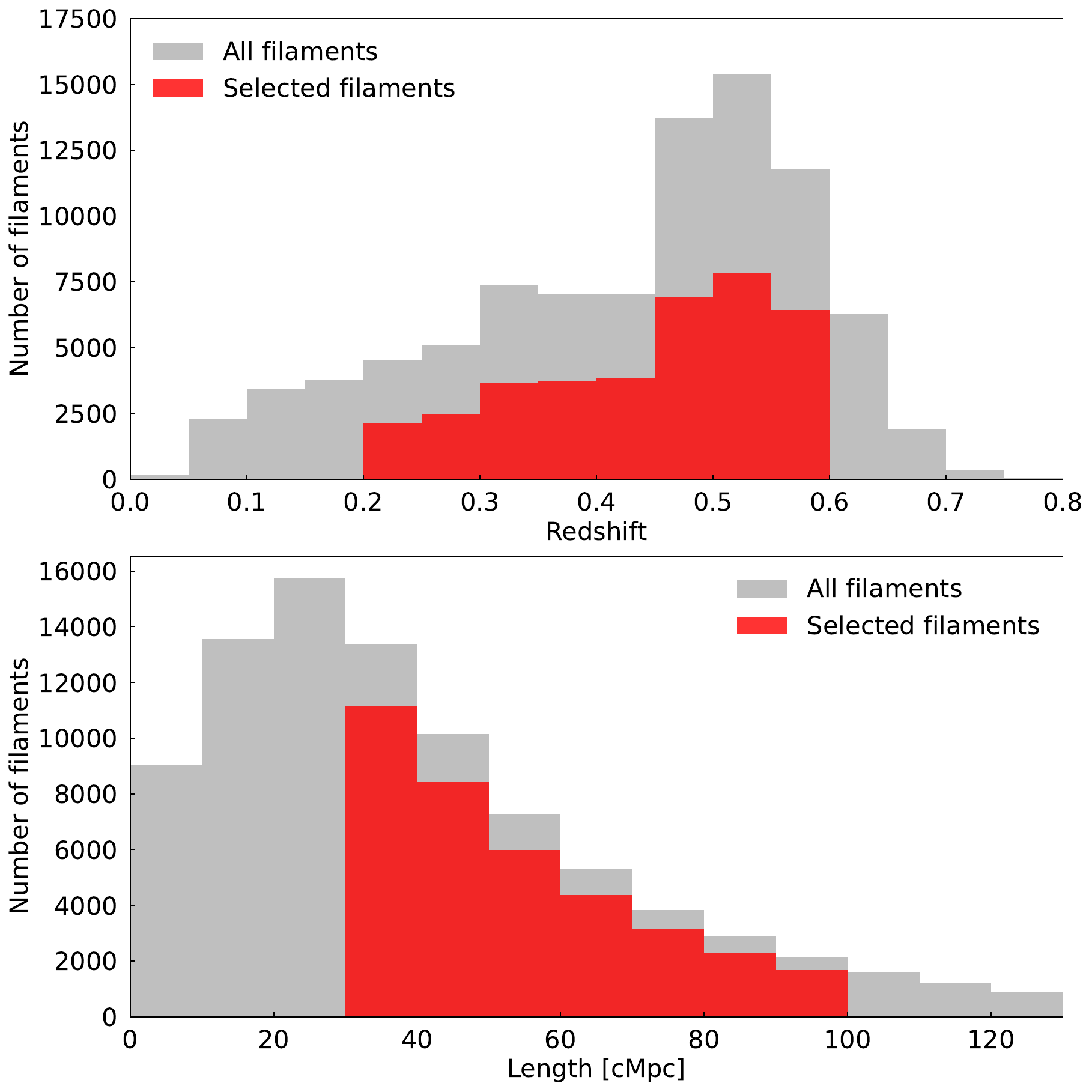}
    \caption{Redshift (top) and length (bottom) distributions of the selected filament sample (highlighted in red). }
    \label{fig:z_l_distribution}
\end{figure}

\begin{figure}[t!]
    \centering
    \includegraphics[width=0.6\textwidth]{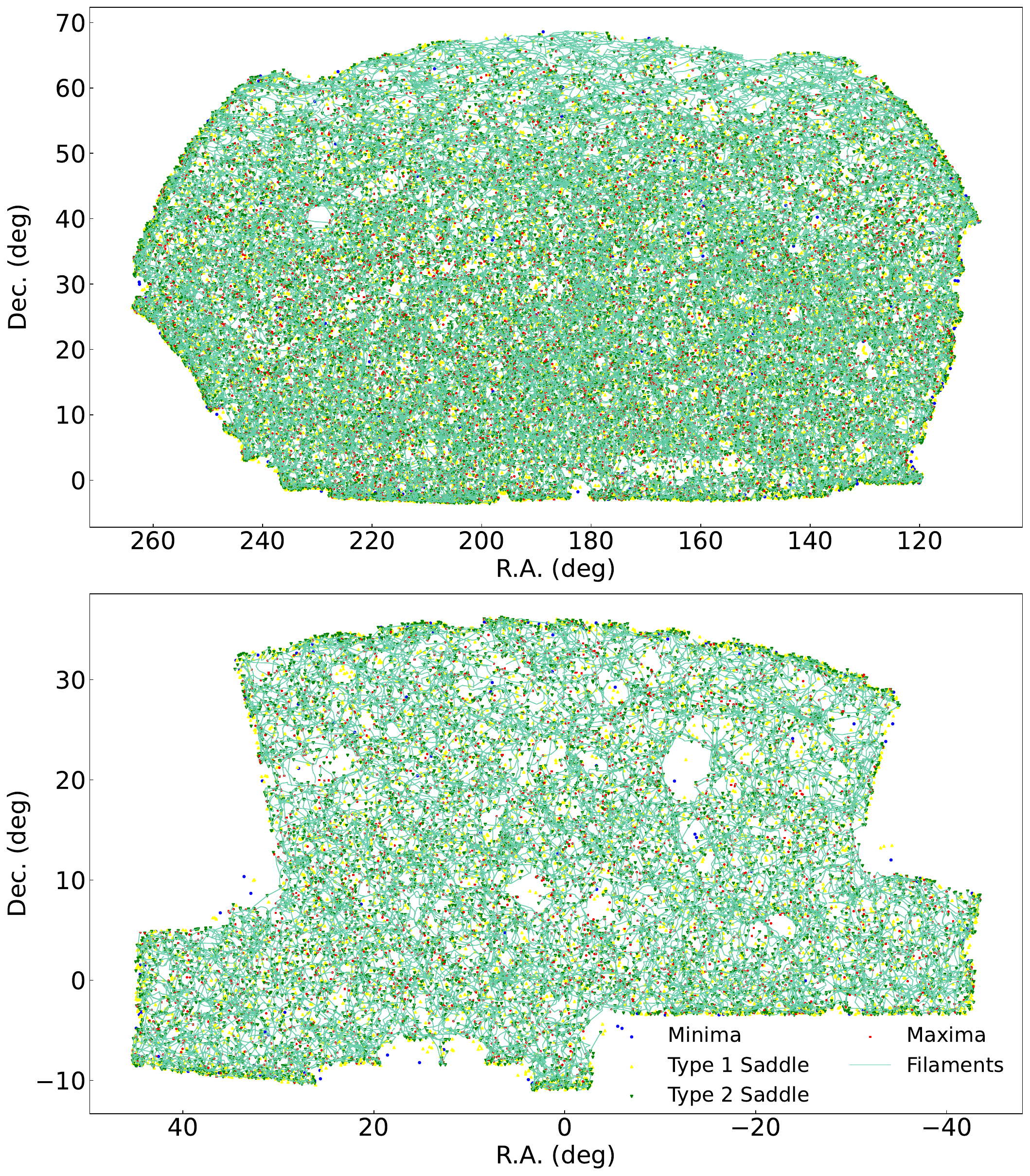}
    \caption{Spatial distribution of filaments extracted by DisPerSE: northern sky (top) and southern sky (bottom).}
    \label{fig:SP_band}
\end{figure}

Our filament catalog is constructed by the DisPerSE algorithm, which is a discrete persistent structure extractor tool suitable for detecting filamentary structures~\cite{Sousbie2011MNRAS.414..350S}. We input SDSS DR12 LOWZ and CMASS\footnote{\url{https://data.sdss.org/sas/dr12/boss/lss/}} galaxies distribution data, which involves 953,193 galaxies in the northern sky and 372,601 galaxies in the southern sky~\cite{Reid2016MNRAS.455.1553R}.

DisPerSE utilizes the Delaunay Tessellation Field Estimator (DTFE) to calculate the density field. Delaunay tessellation divides neighboring galaxies into tetrahedra, with the galaxies serving as the vertices of these tetrahedra~\cite{Schaap2000A&A...363L..29S,vandeWeygaert2009}. As a result, the sky is covered by tetrahedra, where the volume of each tetrahedron reflects the number density of galaxies. \jz{To mitigate boundary effects, DisPerSE provides several boundary treatment options (e.g., \texttt{mirror}, \texttt{smooth}, \texttt{periodic}), controlled by the parameter \texttt{btype}. In this work, we adopt the \texttt{smooth} option, which generates a set of guard particles near the survey edges with a random distribution following the interpolated density field. This approach ensures a smooth transition of the density field across the survey boundaries and suppresses boundary-induced artifacts arising from abrupt density truncation.}
\par
DisPerSE then applies Morse Complex theory to analyze the gradient of the density field. Critical points, where the gradient is zero, are identified. These critical points include maxima, representing high-density peaks, minima, corresponding to voids, and two types of saddle points that represent local density minima bounded by structures, such as walls or filaments. DisPerSE connects these critical points based on the topology of the density field to extract the filamentary structures that form part of the cosmic web~\cite{Sousbie2011MNRAS.414..350S,SousbiePichon2011MNRAS.414..384S}. Filaments are defined as field lines originating from saddle points and connecting maxima, effectively outlining the skeleton of the cosmic web. When multiple filaments intersect or diverge, their arcs can overlap spatially, forming bifurcation points. DisPerSE provides the \texttt{-breakdown} option to identify these bifurcation points and to decompose overlapping filaments into distinct segments.
\par
DisPerSE also provides an option to smooth the density field by iteratively averaging each galaxy’s density with that of its neighbors, defined through the edges of the Delaunay tessellation. This approach effectively suppresses small-scale noise while preserving the underlying large-scale structure. Moreover, DisPerSE  utilizes  persistence theory to exclude unreliable filaments. Persistence is determined by the density contrast between critical points. Filaments with low persistence values, often caused by small variations, may arise from Poisson noise rather than representing real structures. By applying an appropriate persistence threshold, unreliable filaments can be effectively filtered out.
In our analysis, we use the version without smoothing and a $3\sigma$ persistence threshold, ensuring consistency with Malavasi et al.’s results in~\cite{Malavasi2020A&A...642A..19M}. This threshold effectively guarantees that the majority of the identified filaments are real. Finally, a total of 64,110 filaments in the northern sky and 26,121 filaments in the southern sky were extracted by DisPerSE.

\jz{The filamentary network analyzed in this work is extracted in three-dimensional redshift space, where redshift-space distortions (RSD) influence filament identification. In particular, the Fingers-of-God (FoG) effect elongates the galaxy distribution along the line-of-sight (LOS) around clusters, which can be spuriously identified as filaments aligned with the LOS. To mitigate this, we have excluded clusters (masked out to $3 \times R_{500}$) and omitted filaments oriented close to the LOS during the fitting process (see Section~\ref{subsec:proj_bias_corr}). To further assess any residual impact of the FoG effect on our stacking results, we performed a robustness test by applying three alternative treatments to the galaxy distributions around clusters: (1) replacing galaxies within the FoG region with a single cluster-centered object; (2) redistributing the affected galaxies according to an NFW radial profile; and (3) compressing only the LOS-elongated galaxies onto a spherical surface. These methods represent different levels of compression in a physically motivated FoG correction (see Appendix~\ref{app:fog} for details). A comparison of the resulting stacked profiles shows that they are sensitive to the compression level, yet remain mutually consistent within $1\sigma$ uncertainties, except for the most extreme `cluster center replacement' case. This indicates that the FoG effect does not significantly bias our main results.}

Subsequently, following the criteria outlined by Tanimura et al.~(2020)~\cite{Tanimura2020A&A...637A..41T}, filaments were selected within a redshift range of $ 0.2 \sim 0.6 $ and a length range of $30 \sim 100\,\mathrm{cMpc}$, excluding unreliable filaments and those with significant errors. According to Tanimura et al. (2020)’s description~\cite{Tanimura2020A&A...637A..41T,Yang2005MNRAS.357..608Y},
\jz{filaments longer than $100\,\mathrm{cMpc}$ are likely non-physical. This is consistent with the MTNG results of Galárraga-Espinosa et al.\ (2023)~\cite{Daniela2024A&A...684A..63G}, where the filament length distribution shows a sharp cutoff below $100\,\mathrm{cMpc}$ with virtually no structures extending beyond this scale.} In contrast, short filaments ($<30\,\mathrm{cMpc}$) often represent connections between low-mass halos embedded within larger filaments. Therefore, our analysis focuses on filaments longer than $30\,\mathrm{cMpc}$ but shorter than $100\,\mathrm{cMpc}$ to detect and estimate the baryons contained by them. After applying the fiducial redshift and length criteria, a total of $26{,}713$ filaments were identified in the northern sky and $10{,}341$ in the southern sky.

Figure~\ref{fig:z_l_distribution} illustrates the length and redshift distributions of the filament catalog. The gray color represents all 90,231 filaments across the entire sky, while the red color highlights the 37,054 selected filaments. Figure~\ref{fig:SP_band} shows the spatial distribution of selected filaments. The upper panel represents the northern sky, and the lower panel represents the southern sky. The arcs colored in medium aquamarine indicate filaments, while the maxima and minima are marked as red and blue dots. Triangles represent saddle points.

\subsection{Cluster catalogs and masking strategy}
\label{sec:clu_catalog}

\begin{figure}[t!]
    \centering
    \begin{subfigure}{0.6\textwidth}
        \centering
        \includegraphics[width=\linewidth]{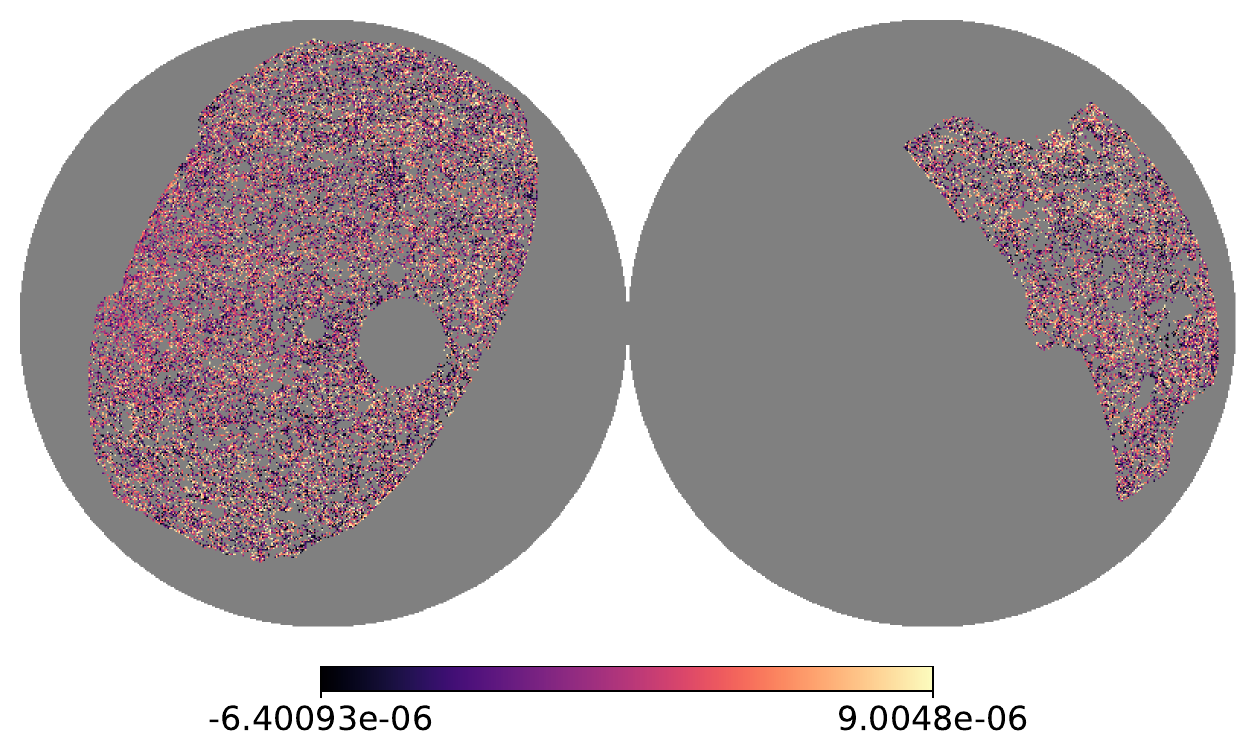}
        \caption{Masked Compton-$y$ map}
    \end{subfigure}
    
    \vspace{6pt}

    \begin{subfigure}{0.6\textwidth}
        \centering
        \includegraphics[width=\linewidth]{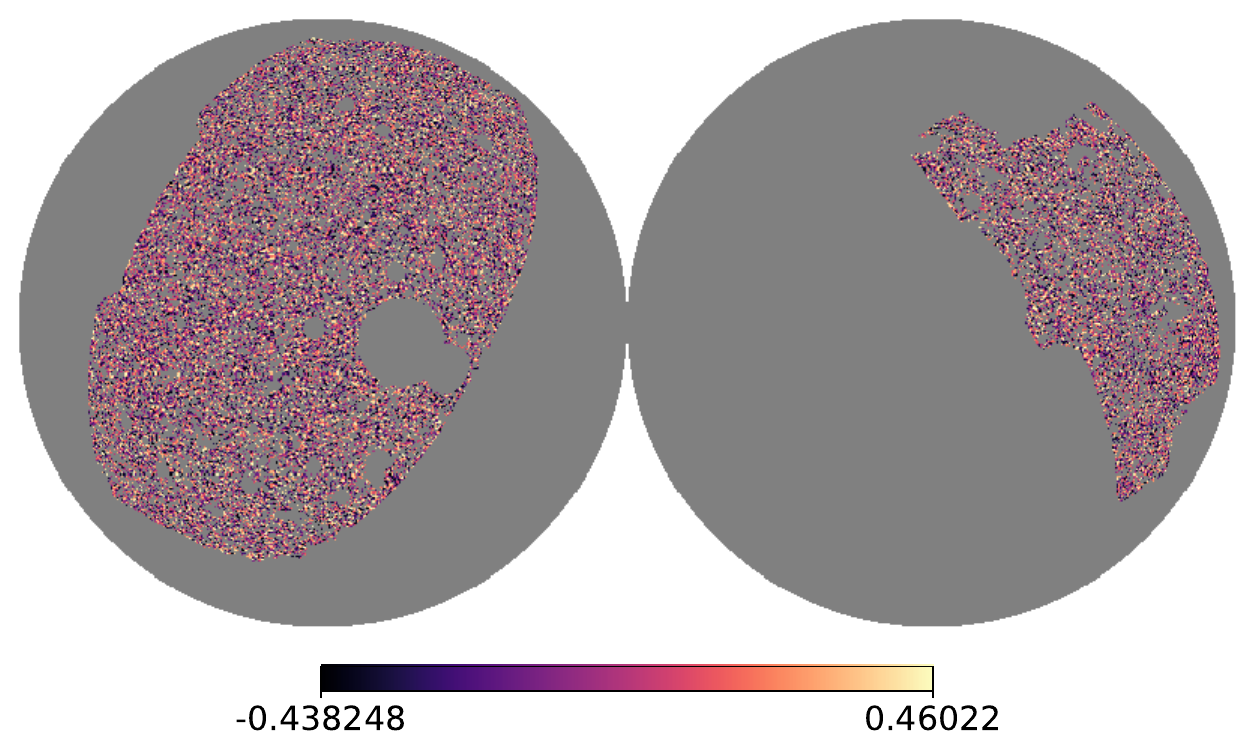}
        \caption{Masked lensing convergence map}
    \end{subfigure}

    \caption{
Masked \textit{Planck} Compton-$y$ \textbf{(a)} and lensing convergence \textbf{(b)} maps centered on the north and south Galactic poles. Regions outside the SDSS DR12 footprint are excluded, and additional masking is applied to remove galaxy clusters out to $3 \times R_{500}$.
}
    \label{fig:masked_maps}
\end{figure}

As a significant fraction of hot gas resides in galaxy clusters, their strong tSZ and gravitational lensing signals can severely contaminate the signal from filaments that we aim to extract. Therefore, removing the cluster contributions is essential for isolating the diffuse WHIM component associated with filaments. 
\par
Following the masking procedure described in Tanimura et al. (2019b)~\cite{Tanimura2019bA&A...625A..67T}, we apply circular masks around the positions of known clusters, including 1,653 \textit{Planck} SZ clusters~\cite{PlanckCollaboration_Ade_Aghanim_Armitage-Caplan_Arnaud_Ashdown_Atrio-Barandela_Aumont_Baccigalupi_Banday_etal._2014}, 1,743 MCXC X-ray clusters~\cite{Piffaretti_Arnaud_Pratt_Pointecouteau_Melin_2011}, 26,111 redMaPPer clusters~\cite{Rykoff_Rozo_Busha_Cunha_Finoguenov_Evrard_Hao_Koester_Leauthaud_Nord_etal._2014}, 158,103 WHL clusters ~\cite{Wen_Han_Liu_2012,Wen_Han_2015}, and 46,479 AMF clusters~\cite{Banerjee_Szabo_Pierpaoli_Franco_Ortiz_Oramas_Tornello_2018}. The mask radii are scaled according to the estimated $R_{500}$ values of each cluster to ensure their full exclusion.
\par
Figure~\ref{fig:masked_maps} shows the Compton-\textit{y} and lensing convergence maps after applying the standard \textit{Planck}-provided masks to exclude point sources and the Galactic plane, along with additional masking of clusters out to $3 \times R_{500}$. Regions outside the SDSS DR12 footprint are also masked, resulting in a left sky region of 24.85\%. These procedures ensure minimal contamination from foreground structures in the subsequent stacking analysis.

\section{Method}
\label{sec:Method}

This section describes the methodology used to stack the tSZ and CMB lensing signals around cosmic filaments. We present the resulting profiles, validate their robustness through bootstrap uncertainties and convergence tests, and apply a bias correction to the observed signals in preparation for the model fitting in the next section.

\subsection{Stacking procedure}

\par
\begin{figure}[t!]
    \centering
    \includegraphics[width=0.6\textwidth]{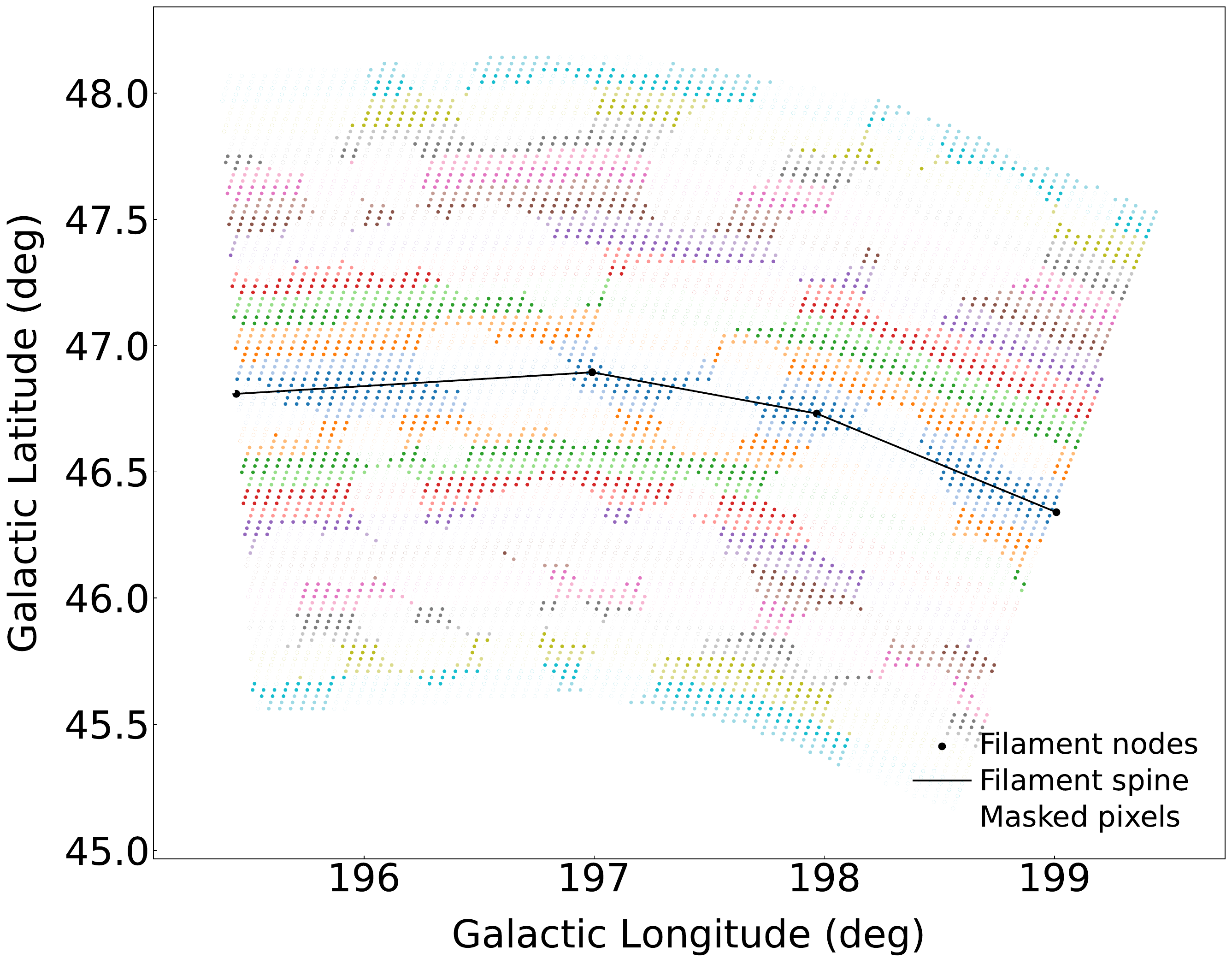}
    \caption{
Schematic illustration of the filament stacking procedure. Black lines represent the filament spine constructed from critical points (black dots) identified by DisPerSE. \jz{The filament endpoints correspond to maxima or bifurcation points, whereas the intermediate nodes correspond to type–1 saddle points along the filamentary structure.} Radial bins (colored) extend perpendicularly from the spine out to $25\,\mathrm{cMpc}$ in comoving radius. Colored dots mark the central positions of HEALPix pixels included in the analysis, while white dots indicate masked pixels excluded from the stacking.
    }
    \label{fig:stack_sketch}
\end{figure}

 
After applying the filament selection criteria and masking the maps, we proceed to stack the tSZ and CMB lensing convergence signals around the selected filament sample. As illustrated in Figure~\ref{fig:stack_sketch}, each filament typically consists of a set of connected segments. The filament spine is traced by black lines connecting critical points (black dots) identified by the DisPerSE algorithm.

We define a region around each filament spine extending out to $25\,\mathrm{cMpc}$ in comoving radius. This region is divided into 20 equally spaced radial bins. For each bin, we identify the corresponding HEALPix pixels in the Compton-$y$ and lensing $\kappa$ maps. The central coordinates of the selected pixels are visualized in color, while masked pixels (e.g., due to point sources, Galactic foregrounds, or galaxy clusters) are shown in white and are excluded from further calculations.

To accurately account for the geometry near junctions between connected filament segments, we explicitly supplement the stacking region with fan-shaped sectors at each junction point. These sectors are constructed based on the angle between adjacent segments and ensure full angular coverage of the filament surroundings. Additionally, to prevent the same pixel from being counted multiple times in overlapping regions (particularly at segment junctions), we assign each HEALPix pixel to the nearest radial bin using a minimum-bin-index rule. This guarantees that each pixel contributes to the stacked profile only once.

\par 
To account for the varying masking fraction across filaments, we assign a weight to each filament based on the fraction of unmasked pixels within the $0\text{--}25\,\mathrm{cMpc}$ radial range. Specifically, the weight $w_i$ is defined as:

\begin{equation}
w_i = \frac{N_{i,\mathrm{valid~pixels}}}{N_{i,\mathrm{total~pixels}}}\,.
\label{eq:weight}
\end{equation}
Filaments that contain at least one radial bin with no valid pixels are excluded from the stacking to avoid introducing noise and bias. 
After applying this criterion with a cluster masking radius of $3 \times R_{500}$, we retain a total of 33,874 filaments in the Compton-$y$ map, comprising 24,342 in the northern sky and 9,532 in the southern sky, and 33,918 filaments in the lensing map, comprising 24,359 in the north and 9,559 in the south, respectively, for the final stacking analysis.

\par 
To mitigate contamination from large-scale background fluctuations, we subtract the mean signal measured in the outer 15--25~cMpc region from each filament profile. This background subtraction is applied individually to both the $y$ and $\kappa$ profiles before stacking. The final stacked signal profiles are computed as the weighted mean of the background-subtracted values across all filaments:

\begin{equation}
\overline{y}(r) = \frac{\sum w_i \left( y_i(r) - y_{i,\mathrm{bg}} \right)}{\sum w_i}\,,
\end{equation}

\begin{equation}
\overline{\kappa}(r) = \frac{\sum w_i \left( \kappa_i(r) - \kappa_{i,\mathrm{bg}} \right)}{\sum w_i}\,,
\end{equation}
where, $y_i(r)$ and $\kappa_i(r)$ denote the Compton-$y$ and lensing convergence profiles measured from the $i$-th filament, while $y_{i,\mathrm{bg}}$ and $\kappa_{i,\mathrm{bg}}$ are the corresponding background values averaged over the outer region ($15$--$25$~cMpc). The weight $w_i$ quantifies the unmasked pixel fraction for each filament, as defined in eq.~(\ref{eq:weight}).

To assess the impact of background selection on the stacked profiles, we also performed an alternative analysis using a background substraction range of $10$--$20\,\mathrm{cMpc}$. The resulting $y$ and $\kappa$ profiles, as well as the inferred model parameters, are consistent with those from the fiducial $15$--$25\,\mathrm{cMpc}$ background choice, indicating the robustness of our signal extraction. We present these results in Appendix~\ref{app:bg}.

\subsection{Stacking results and validation}
\begin{figure}[t!]
    \centering
    \begin{subfigure}[t]{0.49\textwidth}
        \centering
        \includegraphics[width=\textwidth]{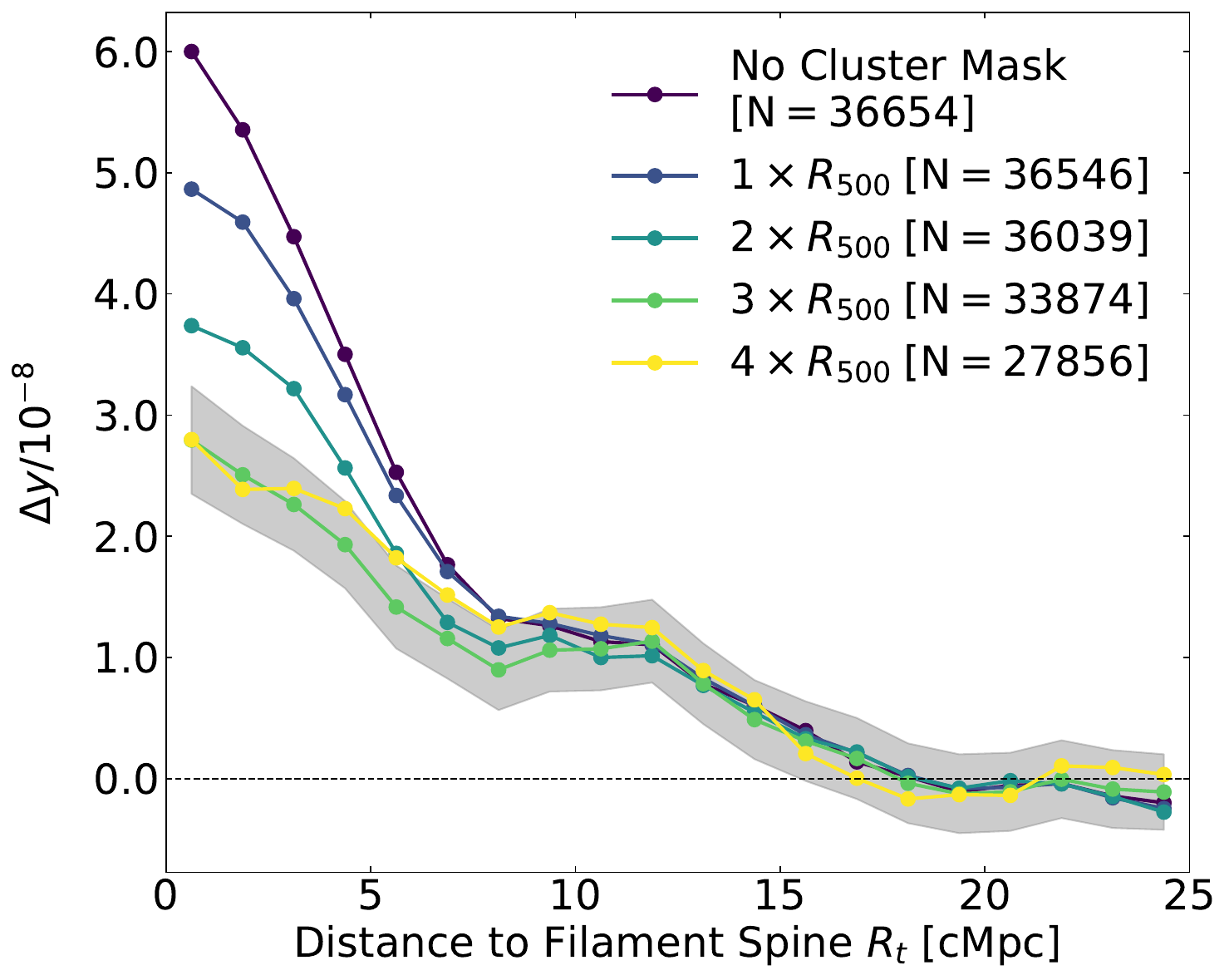}
        \caption{}
        \label{fig:y_profile}
    \end{subfigure}
    \hfill
    \begin{subfigure}[t]{0.49\textwidth}
        \centering
        \includegraphics[width=\textwidth]{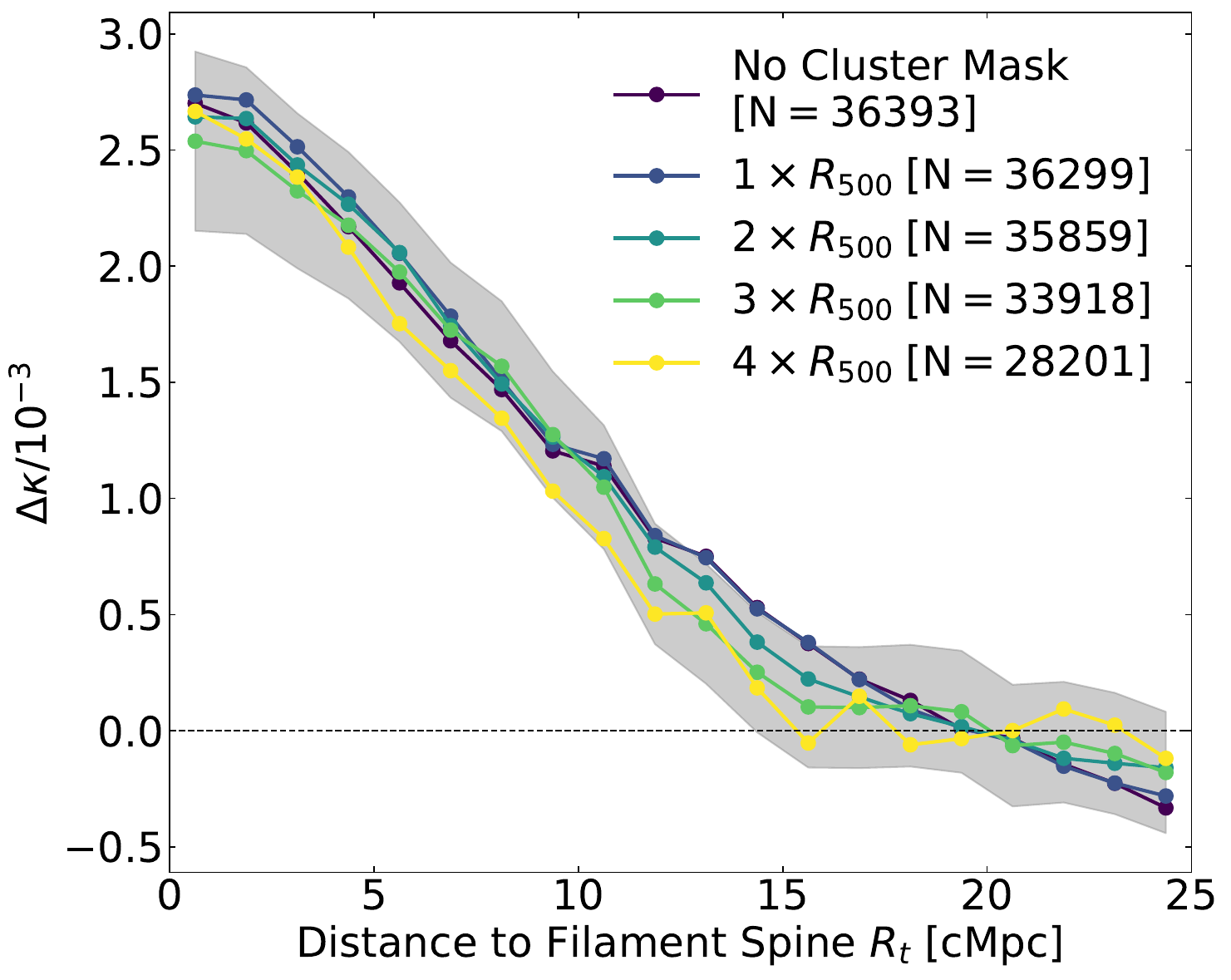}
        \caption{}
        \label{fig:kappa_profile}
    \end{subfigure}
    \caption{
    Average radial profiles of stacked filaments for Compton-$y$ \textbf{(a)} and lensing convergence $\kappa$ \textbf{(b)} under different cluster masking radii. Shaded bands indicate $1\sigma$ bootstrap uncertainties. }
    \label{fig:y_kappa_profiles}
\end{figure}

\par 
We perform a stacking analysis of both the Compton-$y$ and CMB lensing convergence ($\kappa$) signals around filament spines, following the methodology described in the previous section. To evaluate the robustness of the stacked signals against contamination from galaxy clusters, we test a series of masking schemes with increasing radii around known clusters.

We find that the stacked signal profiles converge as the masking radius around known clusters increases, indicating that cluster contributions are effectively mitigated beyond a certain threshold. Based on this convergence behavior, we adopt a final masking radius of $3 \times R_{500}$, consistent with previous studies~\cite{Tanimura2020A&A...637A..41T}. Figure~\ref{fig:y_profile} presents the resulting average Compton-$y$ profiles under different cluster masking radii. A systematic suppression of the $y$ signal is observed in the core region of the filament with increasing mask size, demonstrating the effectiveness of our masking strategy in removing contamination from massive clusters.

The stacked lensing convergence profiles are shown in Figure~\ref{fig:kappa_profile}. From no cluster masking up to a radius of $4 \times R_{500}$, the profiles remain nearly consistent, indicating that the measurements have converged and that contamination from massive clusters is already well controlled within this range. Compared to the Compton-$y$ signal, the lensing convergence profiles exhibit reduced sensitivity to cluster contamination.


The gray error bars in Figures~\ref{fig:y_kappa_profiles} represent the $1\sigma$ uncertainties derived from 1,000 bootstrap realizations of the filament sample, and reflect the statistical variation in the stacked profiles. We quantify the detection significance using a signal-to-noise ratio defined as:
\begin{equation}
\mathrm{S/N} = \sqrt{\chi^2_{\mathrm{data}}}\,,
\end{equation}
where
\begin{equation}
\chi^2_{\mathrm{data}} = \sum_{i,j} s(R_i) \, C^{-1}_{ij} \, s(R_j)\,.
\end{equation}

Here, $s(R_i)$ denotes the stacked signal value (either the Compton-$y$ or the CMB-lensing convergence $\kappa$) at the radial bin $R_i$, and $C^{-1}_{ij}$ is the inverse of the bootstrap-derived covariance matrix. \jz{Using this definition, we evaluate the significance of the measured profiles with respect to the null (zero-signal) hypothesis over the radial range $0$--$15\,\mathrm{cMpc}$ for the $3 \times R_{500}$ masking case, corresponding to the green curve in Fig.~\ref{fig:y_kappa_profiles}.} We detect the tSZ signal at a significance of $7.82\sigma$ and the CMB-lensing signal at $7.78\sigma$.

\jz{To test the robustness of our results and explore potential dependencies, we perform additional stacking analyses by dividing the filament sample into redshift, length, and environmental overdensity bins. The detailed results are presented in Appendix~\ref{app:subsamples}, Fig.~\ref{fig:subsample_yk_profiles}. We find a marginally enhanced $y$-signal for higher-redshift filaments ($0.47 < z < 0.6$), which appears to drive the bump feature around $10\,\mathrm{cMpc}$. This feature may originate from the contribution of nearby filaments that fall within the stacking aperture at this separation scale.
Regarding the dependence on filament length, we observe that shorter filaments exhibit a stronger tSZ signal. This trend may reflect the fact that very long filaments tend to include merged or noisy segments, which dilute the average stacked signal. Finally, we find a modest enhancement in both the tSZ and CMB-lensing signals for filaments residing in high-density environments, consistent with expectations from denser gas contributions.}

\subsection{Projection bias correction}
\label{subsec:proj_bias_corr}
When fitting the stacked profiles with an analytical filament model, we assume that filaments are infinitely long, straight, and oriented perpendicular to the LOS. In reality, filaments have finite lengths and exhibit a broad distribution of inclination angles relative to the plane of the sky, which introduces geometric projection effects in the LOS integration. \jz{This projection contamination is most significant near filament endpoints, where the integration cylinder extends beyond the physical filament, thereby suppressing the observed signal (see Fig. \ref{fig:bias_diagram} in  Appendix~\ref{app:geometry_bias}).}

\jz{It is possible to incorporate orientation information into the modeling framework, as demonstrated in the analytical approach of~\cite{Tanimura2020A&A...637A..41T}. However, extending the simple model for perpendicular filaments described in Section~\ref{sec:model} would necessitate additional assumptions regarding the 3D orientation distribution and projection geometry, and would incur a higher computational cost. We therefore adopt a data-driven correction method in this work.}

\begin{figure}[t!]
    \centering
    \begin{minipage}[t]{0.49\textwidth}
        \centering
        \includegraphics[width=\textwidth]{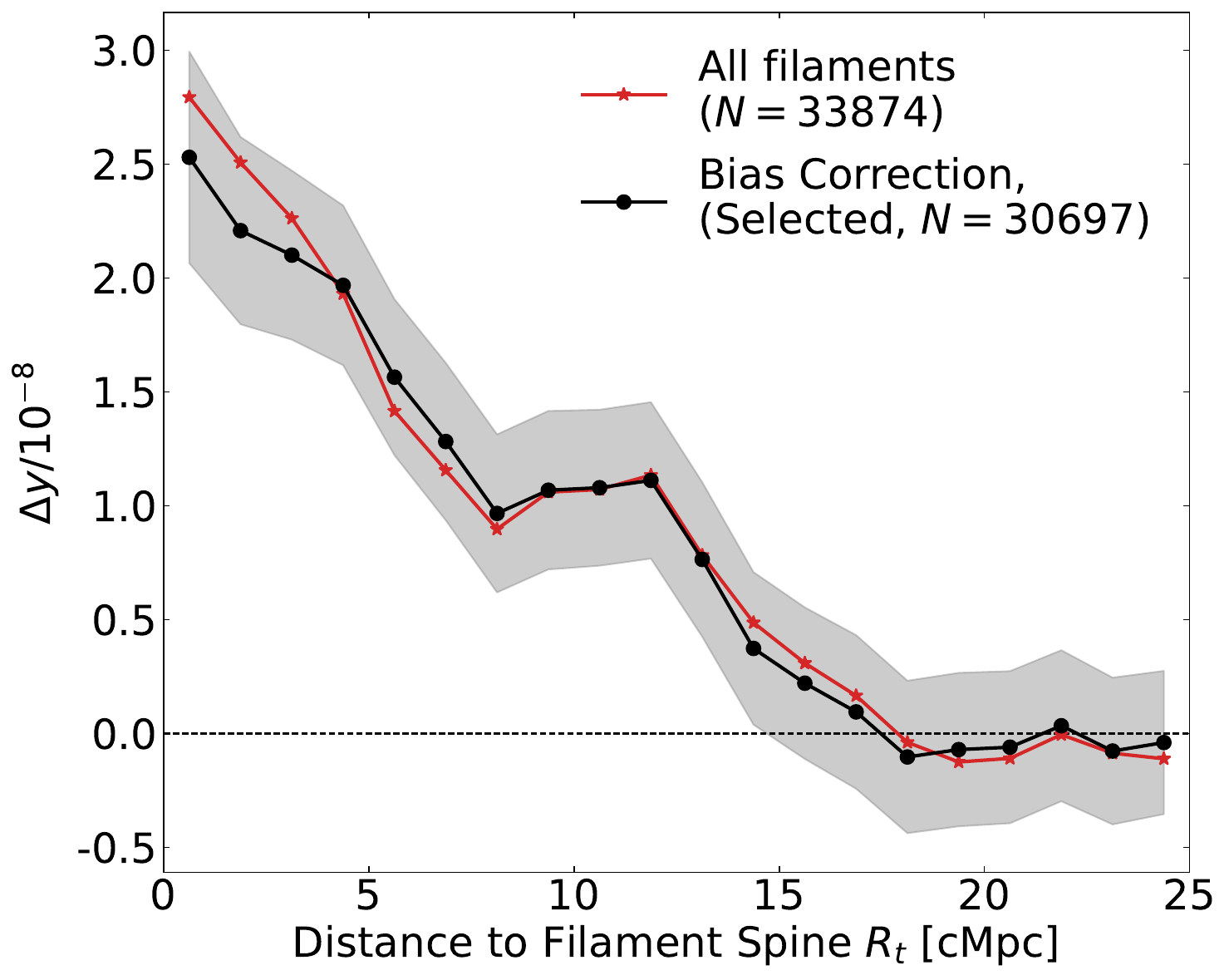}
    \end{minipage}\hfill
    \begin{minipage}[t]{0.49\textwidth}
        \centering
        \includegraphics[width=\textwidth]{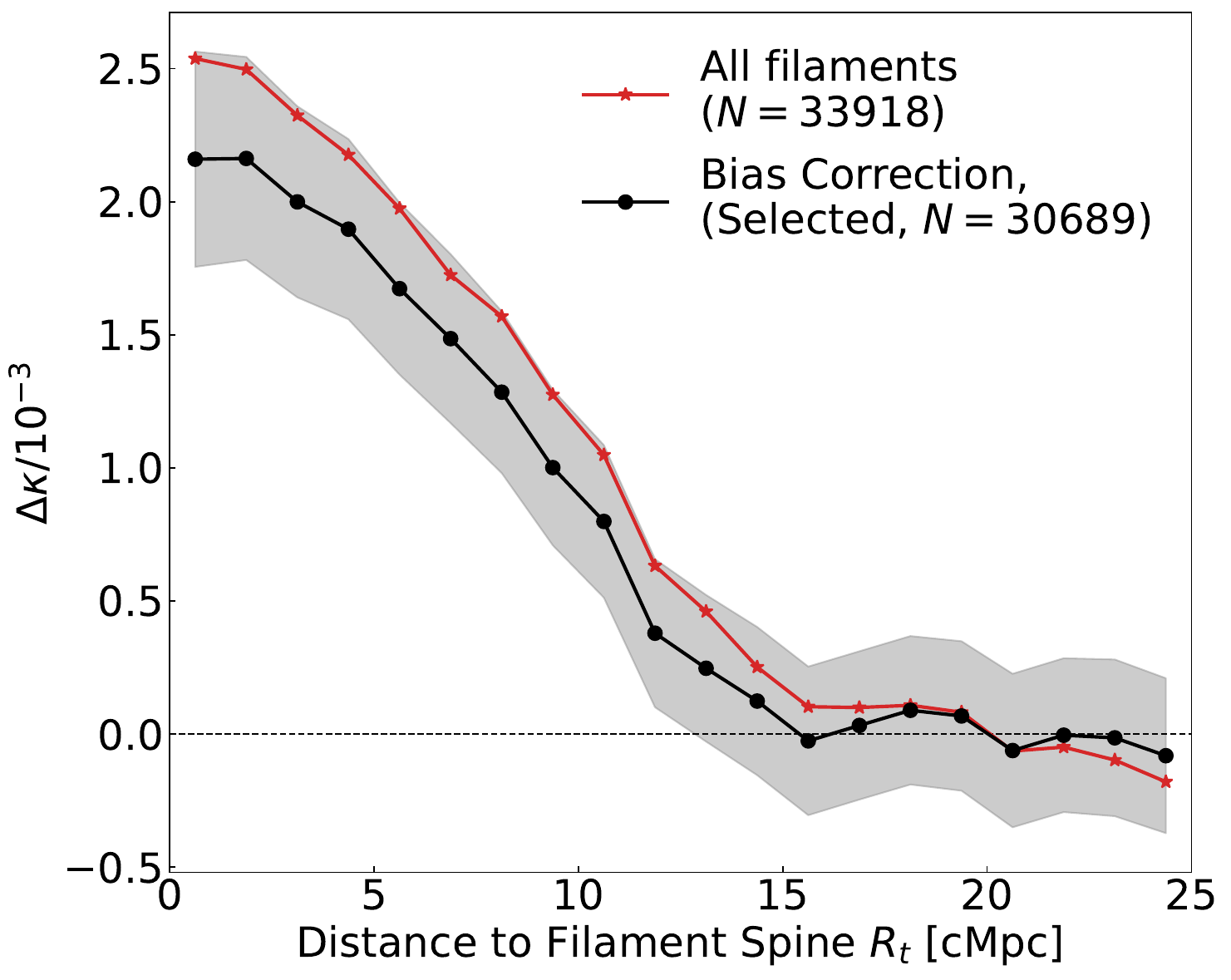}
    \end{minipage}
    \caption{
Correction of geometric projection bias in the stacked radial profiles.  
\textbf{Left}: Compton-$y$ profiles.  
\textbf{Right}: Lensing convergence ($\kappa$) profiles. In both panels, the red curve shows the original (uncorrected) signal. The black curve shows the corrected signal after applying the bias correction described in eq.~\eqref{eq:signal_correction}, computed over a filtered filament subset that excludes filaments with $\tan \theta < 25 / L$ to mitigate extreme projection contamination.
}
    \label{correction}
\end{figure}

\jz{To account for the projection effects, we estimate for each filament the fractional bias induced by its inclination angle $\theta$ and length $L$, and apply the geometric correction factor defined in Eq.~\eqref{eq:bias_formula}. Filaments with extreme inclination angles ($\tan \theta < 25 / L$), where the projection contamination becomes severe, are removed from the sample. After applying this selection, approximately 30,700 filaments remain for profile modeling and parameter fitting. Figure~\ref{correction} illustrates the effect of applying the inclination and geometric corrections on the stacked Compton-$y$ and lensing convergence $\kappa$ profiles. Further details of the projection-bias correction are provided in Appendix~\ref{app:geometry_bias}.}

\section{Model}
\label{sec:model}

In this section, we model the tSZ (Compton-$y$) and CMB lensing convergence ($\kappa$) signals around cosmic filaments and apply resolution-matching procedures to ensure consistency with the observational data. These models serve as the basis for the joint fitting analysis presented in the next section.
\subsection{Modeling the tSZ signal}
\label{sec:ymodel}

The tSZ effect arises from inverse Compton scattering of CMB photons off high-energy electrons in ionized gas. As CMB photons travel from the surface of last scattering to the present day, they may encounter regions of hot, ionized gas. In such regions, scattering boosts the average photon energy, resulting in a spectral distortion of the CMB blackbody spectrum~\cite{Zeldovich1969Ap&SS...4..301Z}. The tSZ signal, typically expressed in terms of the Compton-$y$ parameter, is given by:

\begin{equation}
y(\hat{\mathbf{n}}) = \frac{\sigma_Tk_B }{m_e c^2} \int \frac{d\chi}{1+z} \, n_e(\chi \hat{\mathbf{n}}) \, T_e(\chi \hat{\mathbf{n}})\,,
\label{eq:compton_y}
\end{equation}
where $\sigma_T$ is the Thomson cross-section, $m_e$ is the electron mass, $c$ is the speed of light, $n_e$ is the phyiscal electron number density, $T_e$ is the electron temperature, and $\chi$ is the comoving distance to redshift $z$.

We model the stacked $y$-signal around filaments by assuming a cylindrically symmetric, isothermal gas distribution. For a filament orienting perpendicular to the LOS, the observed $y$-profile as a function of projected radius $R_t$ (i.e., the tangential distance from the filament spine on the sky) is expressed as:

\begin{equation}
y(R_t, z) = \frac{\sigma_{\mathrm{T}} \, k_{\mathrm{B}} \, T_e}{m_e c^2} \int_{R_t}^{R_{\mathrm{max}}} \frac{2r \, n_e(r,z)}{\sqrt{r^2 - R_t^2}} \, \frac{dr}{(1+z)}\,.
\label{eq:yprofile}
\end{equation}
Here, the radii $r$, $R_t$, and $R_{\mathrm{max}}$ are all defined in comoving coordinates.
\par
We model the radial electron density profile $n_e(r,z)$ using the  $\beta$-model~\cite{Cavaliere1978A&A....70..677C} of the form:

\begin{equation}
n_e(r, z) = \frac{n_{e,0}(z)}{[1 + (\frac{r}{r_{e,c}})^{\alpha}]^{3\beta/2}}\,, \quad \text{for } r < R_{\mathrm{max}}\,,
\end{equation}
where $n_{e,0}(z)$ is the central electron density at redshift $z$, $r_{c}$ is the core radius of the electron distribution, $R_{\mathrm{max}} = 15\,\mathrm{cMpc}$ is the outer cutoff radius, and $(\alpha, \beta)$ are free shape parameters. In our analysis, we explore two variants of the $\beta$-model commonly used to describe the radial distribution of gas and matter in filamentary structures:

\begin{itemize}
    \item \textbf{Standard $\beta$-model:} This is the standard isothermal $\beta$-model with $(\alpha, \beta) = (2, 2/3)$, corresponding to a flat core and a outer slope $\propto r^{-2}$. It has been widely used in modeling intracluster media, and in modeling the baryons in filaments (e.g.~\cite{Tanimura2020A&A...637A..41T,Tuominen2021A&A...646A.156T,zhu2021ApJ...920....2Z,Galarraga-Espinosa2022A&A...661A.115G}).
    \item \textbf{Generalized $\beta$-model:} A profile with $(\alpha, \beta) = (1, 1)$, which introduces a flatter ($\propto r^{-1.5}$) radial decay at outer regions. This form allows greater flexibility in capturing filamentary gas distributions that may differ from cluster-like environments~\cite{yang2025arXiv250702476Y}.
\end{itemize}

We assume that the central electron overdensity, $\delta_{e,0}$, is independent of redshift. Accordingly, the central electron density at redshift $z$ can be expressed as:
\begin{equation}
n_{e,0}(z) = \left(1 + \delta_{e,0}\right) \, \bar{n}_e(z)\,,
\label{eq:ne0}
\end{equation}
where $\bar{n}_e(z)$ is the mean background electron density at redshift $z$, scaling as $\bar{n}_e(z) = \bar{n}_e(0)(1+z)^3$. Here, $\bar{n}_e(0)$ is the mean electron number density at $z=0$, computed from the baryon density parameter $\Omega_b$ as:

\begin{equation}
\bar{n}_e(0) = \frac{\rho_{b,0}}{\mu_e m_p} = \frac{\Omega_b \rho_{\mathrm{crit},0}}{\mu_e m_p}\,,
\end{equation}
where $\mu_e \approx 1.14$ is the mean molecular weight per free electron, and $m_p$ is the proton mass.

The electron temperature $T_e$ is assumed to be constant across all filaments, and is treated as a free parameter in our fitting process.

\subsection{Modeling the CMB lensing convergence signal}
\label{sec:kappa_model}

The gravitational lensing convergence, $\kappa(R)$, traces the projected matter distribution along the LOS and is computed by integrating the 3D matter overdensity weighted by the cosmological lensing kernel~\cite{Lewis2006PhR...429....1L}.

Since our source plane corresponds to the CMB at redshift $z_{\mathrm{CMB}} \sim 1100$, the convergence must be derived within a fully cosmological framework. In this formalism, $\kappa$ 
is expressed as:

\begin{equation}
\kappa = \int_0^{\chi_\ast} W(\chi) \, \delta_m(\chi\hat{\mathbf{n}}) \, d\chi\,,
\end{equation}
where $\chi_\ast$ is the comoving distance to the last scattering surface, $\delta_m(\chi\hat{\mathbf{n}})$ is the matter overdensity field, and $W(\chi)$ is the CMB lensing kernel:

\begin{equation}
W(\chi) = \frac{3 H_0^2 \Omega_m}{2 c^2} \cdot \frac{\chi}{a(\chi)} \left(1 - \frac{\chi}{\chi_\ast} \right)\,,
\end{equation}
with $a(\chi) = (1 + z)^{-1}$ being the cosmic scale factor, $H_0$ being the present-day Hubble constant, and $\Omega_{m,0}$ being the present-day matter density parameter. This kernel peaks at intermediate redshifts, maximizing the lensing efficiency between the observer and the source.

Following CMB lensing analyses, we apply this framework to model the lensing signal from cosmic filamentary structures. For a cylindrically symmetric filament at redshift $z$ lying perpendicular to the LOS, the 2D convergence profile becomes:
\begin{equation}
\kappa(R_t, z) = W(z) \int_{R_t}^{R_{\mathrm{max}}} \delta_m(r) \frac{2r}{\sqrt{r^2 - R_t^2}} \, dr\,,
\label{eq:kappa_general_overdensity}
\end{equation}
where $\delta_m(r)$ is the radial matter overdensity profile. The lensing kernel $W(z)$ in this expression takes the redshift-resolved form:
\begin{equation}
W(z) = \frac{3 H_0^2 \Omega_{m,0}}{2 c^2} \, (1 + z) \, \chi(z) \left(1 - \frac{\chi(z)}{\chi(z_{\mathrm{CMB}})} \right)\,.
\label{eq:lensing_kernel_z}
\end{equation}


We adopt the two $\beta$-models for the matter overdensity profile, consistent with the approach used for the Compton-$y$ signal. The functional form is given by:

\begin{equation}
\delta_m(r) = \frac{\delta_{m,0}}{\left[ 1 + \left( \frac{r}{r_{m,c}} \right)^{\alpha} \right]^{3\beta/2}}\,, \quad \text{for } r < R_{\mathrm{max}}\,,
\end{equation}
where $\delta_{m,0}$ is the central matter overdensity, $r_{m,c}$ is the core radius, and $(\alpha, \beta)$=$(2,2/3)$ or $(1,1)$ control the intermediate slope and outer decay of the profile. Both models are truncated at $R_{\mathrm{max}} = 15\,\mathrm{cMpc}$, and are used to fit the stacked $\kappa$ profiles in order to evaluate the robustness of our results against the assumed matter distribution.


\subsection{Beam convolution and resolution matching}
\label{sec:biascorrection}
In order to allow direct comparison between the analytical model and the stacked observations, we further apply beam smoothing and filtering to the model profiles to match the resolution of the observational data.

For the tSZ signal, the model $y$-profile is convolved with a Gaussian beam kernel:

\begin{equation}
y_{\mathrm{mod}}(R_t, z) = y(R_t, z) \ast B\,,
\end{equation}
where $B$ denotes a Gaussian kernel with full width at half maximum (FWHM) of $10'$—matching the angular resolution of the \textit{Planck} $y$-map~\cite{Chandran2023MNRAS.526.5682C}.

For the lensing convergence model, we mimic the resolution limitation of the \textit{Planck} CMB lensing map by suppressing the high-frequency components of the profile. Specifically, we apply an exponential damping to the model profile beyond multipole $\ell > 400$, which approximately corresponds to a transverse scale of $r \sim 14~\mathrm{cMpc}$ at $z=0.44$. These convolution and filtering processes ensure that both the $y$ and $\kappa$ model profiles are smoothed consistently with the \textit{Planck} data, allowing unbiased comparison in the fitting process.


\section{Results}
In this section, we present the key findings of our study, including the joint MCMC fitting of the stacked Compton-$y$ and CMB lensing convergence ($\kappa$) profiles, which yield constraints on the spatial distribution and thermodynamic properties of gas within cosmic filaments. We also quantify the contribution of the WHIM associated with these filaments to the cosmic baryon budget through model-based gas mass estimates.
\label{sec:results}
\subsection{MCMC fitting}
We perform a joint fit of the $\beta$-models to the stacked Compton-$y$ and lensing convergence ($\kappa$) profiles using the \texttt{emcee} MCMC sampler~\cite{Foreman2013PASP..125..306F,Goodman2010CAMCS...5...65G}. Following the assumption—supported by hydrodynamic simulations~\cite{Gheller2019MNRAS.486..981G}—that baryons approximately trace the underlying dark matter potential on large scales, independent of stellar and AGN feedback, we adopt a common central overdensity and core radius for both gas and matter components. Specifically, we set $\delta_{e,0} = \delta_{m,0} \equiv \delta$ and $r_{e,c} = r_{m,c} \equiv r_c$, where the subscripts “$e$” and “$m$” refer to the electron (gas) and matter profiles, respectively.

The model thus includes three free parameters: the central overdensity $\delta$, the core radius $r_c$, and the electron temperature $T_e$. We adopt uniform priors for each parameter: $0 < \delta < 100$, $0 < r_c < 7\,\mathrm{cMpc}$ and $10^5\,\mathrm{K} < T_e < 10^7\,\mathrm{K}$.

We evaluate the goodness-of-fit by minimizing the total chi-square:
\begin{equation}
\chi^2 = \chi^2_y + \chi^2_\kappa\,,
\end{equation}
where
\begin{equation}
\chi^2_y = \sum_{i,j} \left[y(R_i) - y_{\mathrm{mod}}(R_i)\right] \, C^{-1}_{y,ij} \, \left[y(R_j) - y_{\mathrm{mod}}(R_j)\right]\,,
\end{equation}
\begin{equation}
\chi^2_\kappa = \sum_{i,j} \left[\kappa(R_i) - \kappa_{\mathrm{mod}}(R_i)\right] \, C^{-1}_{\kappa,ij} \, \left[\kappa(R_j) - \kappa_{\mathrm{mod}}(R_j)\right]\,.
\end{equation}
Here, $y(R_i)$ and $\kappa(R_i)$ are the observed values in the $i$th bin, $y_{\mathrm{mod}}(R_i)$ and $\kappa_{\mathrm{mod}}(R_i)$ are the model predictions, and $C_y$ and $C_\kappa$ denote the covariance matrices of the $y$ and $\kappa$ profiles, respectively, estimated via bootstrap resampling.

\begin{figure}[t!]
    \centering
    \begin{minipage}[t]{0.48\textwidth}
        \centering
        \includegraphics[width=\textwidth]{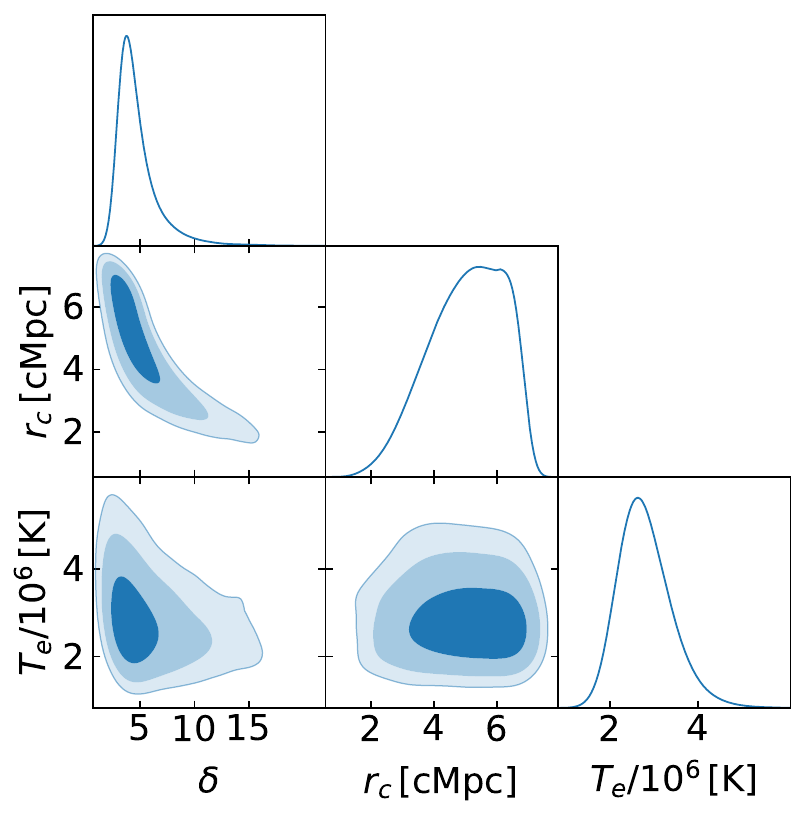}
    \end{minipage}\hfill
    \begin{minipage}[t]{0.48\textwidth}
        \centering
        \includegraphics[width=\textwidth]{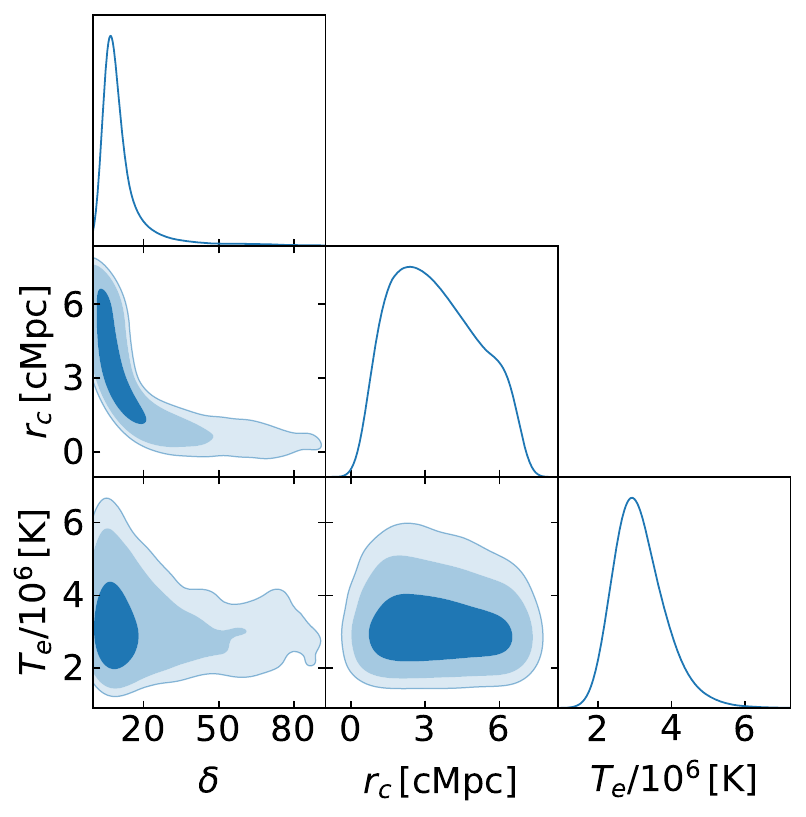}
    \end{minipage}
    \caption{
Corner plots of the posterior distributions for the model parameters from the MCMC fitting of the stacked $y$ and $\kappa$ profiles. 
\textbf{Left}: standard $\beta$-model. 
\textbf{Right}: generalized $\beta$-model.
The diagonal panels show marginalized one-dimensional distributions of central overdensity $\delta$, core radius $r_{c}$, and electron temperature $T_e$.
The off-diagonal panels show the two-dimensional joint posteriors, with contours indicating the $1\sigma$, $2\sigma$, and $3\sigma$ credible regions.
}
    \label{fig:mcmc_corner}
\end{figure}

\begin{figure}[t!]
    \centering
    \begin{minipage}[t]{0.49\textwidth}
        \centering
        \includegraphics[width=\textwidth]{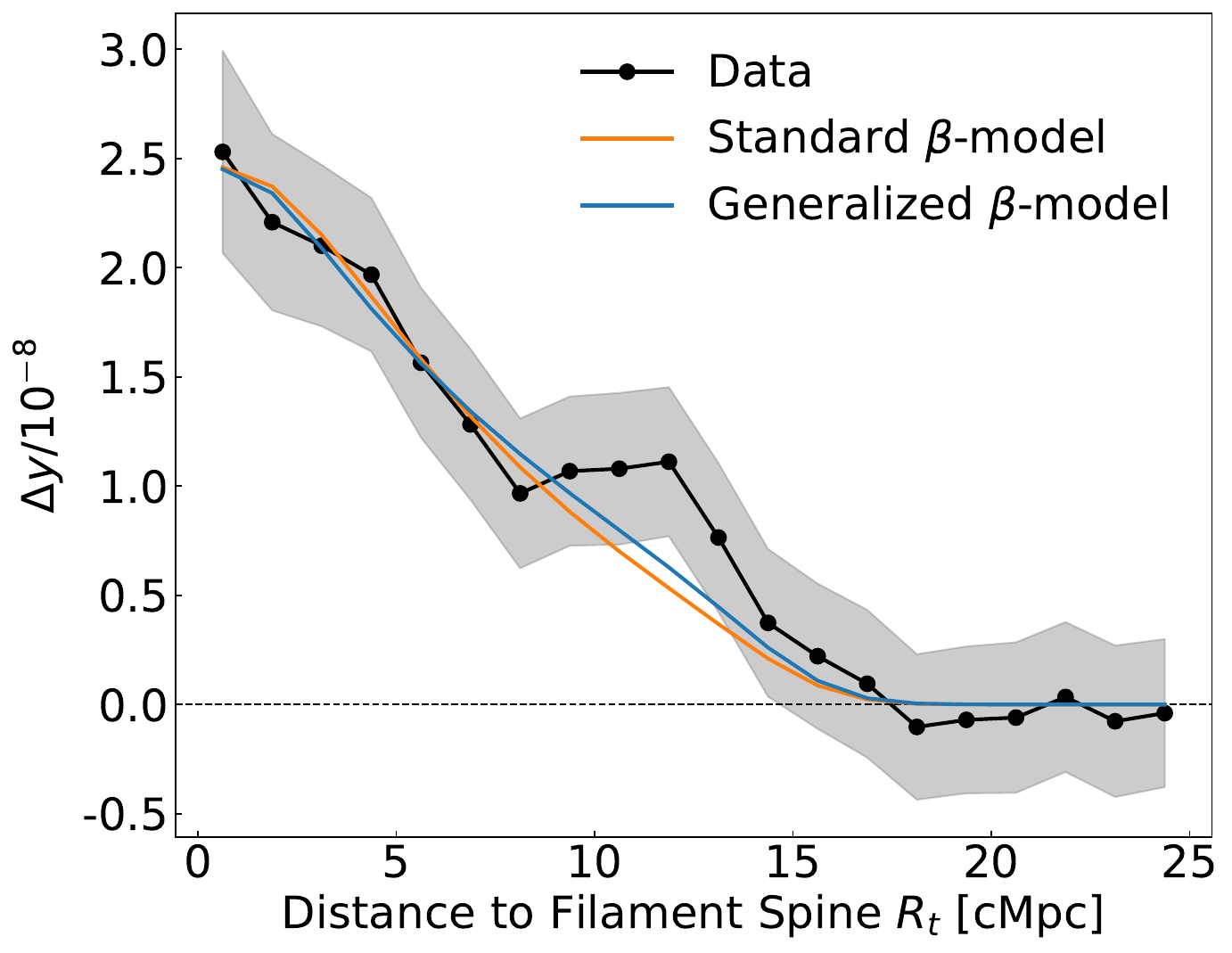}
    \end{minipage}\hfill
    \begin{minipage}[t]{0.49\textwidth}
        \centering
        \includegraphics[width=\textwidth]{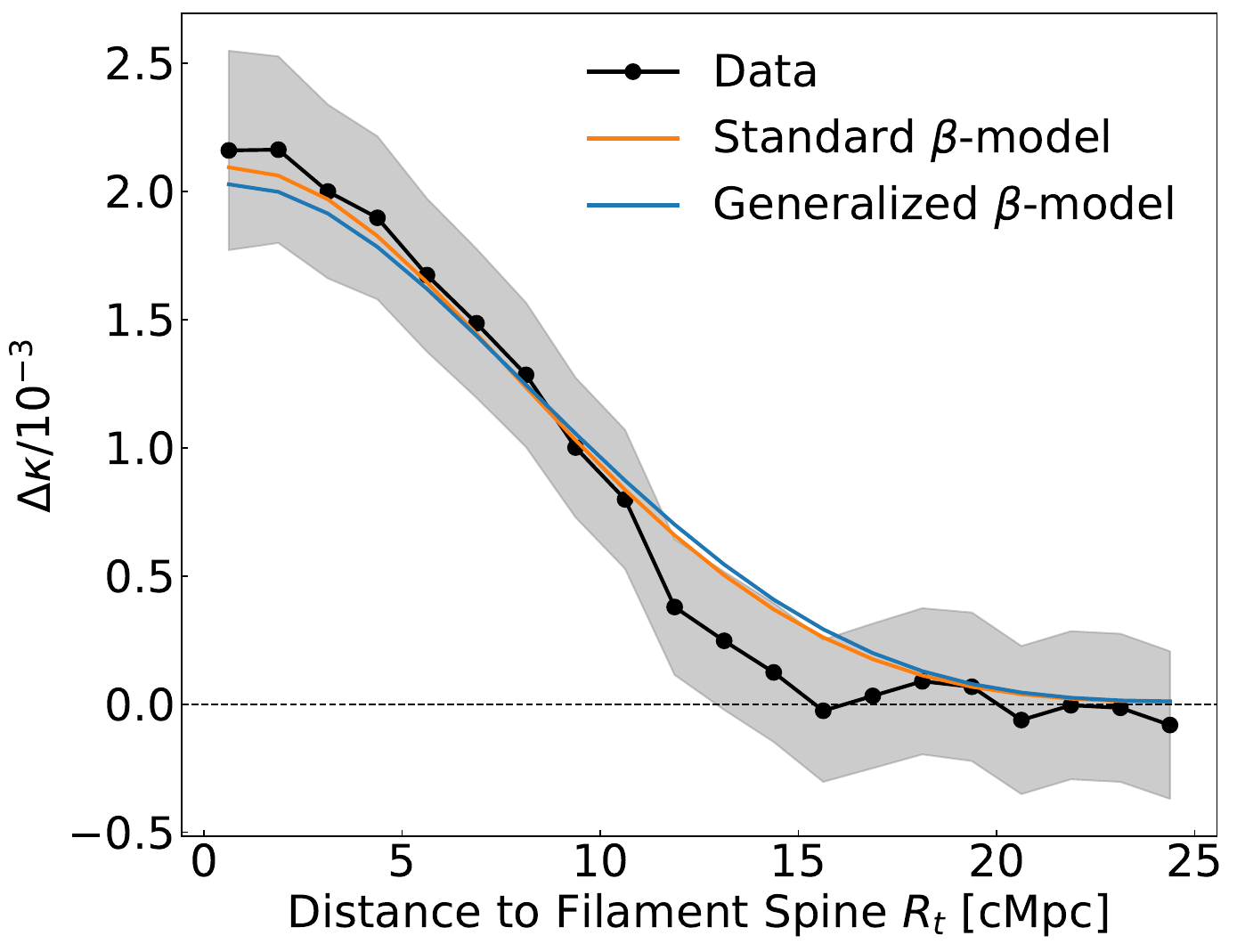}
    \end{minipage}
    \caption{
Average radial profiles of $\sim 30,700$ filaments fitted with the standard $\beta$-model (orange) and generalized $\beta$-model (blue).
\textbf{Left}: Compton-$y$ profiles.
\textbf{Right}: CMB lensing convergence ($\kappa$) profiles.
}
    \label{fig:fitresults}
\end{figure}

The best-fit parameters for the two $\beta$-model parameterizations are summarized in Table~\ref{tab:fit_params}, and the corresponding posterior distributions are shown in Figure~\ref{fig:mcmc_corner}. Figure~\ref{fig:fitresults} shows the model profiles with the best-fit values derived above. The best-fit parameters from the standard $\beta$-model indicate that the filament gas has a modest central overdensity of $\delta \sim 4$. The inferred core radius of $r_c \sim 5~\mathrm{cMpc}$ reflects the broad transverse extent of filaments, consistent with the extended radial trend of the signal observed in our stacked profiles. The electron temperature of $T_e \sim 2.7 \times 10^6~\mathrm{K}$ confirms that the gas resides in the WHIM phase. A strong degeneracy is present between $\delta$ and $r_c$, such that in the generalized $\beta$-model, a smaller core radius is compensated by a higher central overdensity. Owing to the relatively high signal-to-noise ratio of the stacked Compton-$y$ profiles, both model parameterizations yield well-constrained temperature estimates, reinforcing the detection of warm gas in filamentary structures.

\begin{table}[t!]
\renewcommand{\arraystretch}{1.6} 
\centering
\begin{tabular}{
    >{\columncolor{{C0}!20}}l |
    >{\columncolor{{C0}!10}}c |
    >{\columncolor{{C0}!10}}c }
\rowcolor{{C0}!25}
\hline
Parameter & Standard $\beta$-model & Generalized $\beta$-model \\
\hline
$\delta$ 
    & $4.18^{+2.01}_{-1.06}$
    & $7.49^{+8.88}_{-2.87}$ \\
$r_{c}~[\mathrm{cMpc}]$ 
    & $5.18^{+1.20}_{-1.39}$
    & $3.28^{+2.16}_{-1.72}$ \\
$T_e~[10^6~\mathrm{K}]$ 
    & $2.74^{+0.65}_{-0.53}$
    & $3.06^{+0.78}_{-0.60}$ \\
\hline
\end{tabular}
\caption{Best-fit parameters for the standard and generalized $\beta$-models from the joint MCMC fit to the $y$ and $\kappa$ profiles. Uncertainties represent the 16th and 84th percentiles of the marginalized posterior distributions.\label{tab:fit_params}}
\end{table}

\subsection{Baryon budget}
To assess the contribution of filamentary structures to the cosmic baryon budget, we estimate the total gas mass contained in our filament sample based on the $\beta$-models described above. Assuming cylindrical symmetry, the gas mass for an individual filament of comoving length $L$ at redshift $z$ is calculated as:
\begin{equation}
M_{\mathrm{gas}} = \int_0^{L} dl \int_0^{R_{\mathrm{max}}} 2\pi r \rho_{\mathrm{gas}}(r, z)  dr,
\end{equation}
where $\rho_{\mathrm{gas}}(r, z) = \mu_e m_p n_e(r, z)$ is the gas mass density. 

We compute $M_{\mathrm{gas}}$ for each filament in our catalog using its observed redshift and length, and sum over all filaments to obtain the total gas mass. The corresponding average gas mass density is then derived by dividing the total mass by the comoving volume and the sky fraction used in the analysis:
\begin{equation}
{\rho}_{\mathrm{gas}}^{\mathrm{fil}} = \frac{\sum_i M_{\mathrm{gas},\mathrm{fil},i}}{V_c \cdot f_{\mathrm{sky}}},
\end{equation}
and compared to the cosmological mean baryon density $\bar{\rho}_b = \Omega_b \rho_{\mathrm{crit}}(z=0)$.

We estimate the uncertainty on the gas-to-baryon fraction $\rho_{\mathrm{gas}}^{\mathrm{fil}} / \bar{\rho}_b$ using Monte Carlo sampling from the posterior distributions of $\delta_{e,0}$ and $r_{e,c}$ obtained from the joint MCMC analysis of the stacked $y$ and $\kappa$ profiles.

For our analysis, the comoving volume corresponds to the redshift range $0.2 < z < 0.6$, and the effective sky fraction used is $f_{\mathrm{sky}} = 0.2485$, based on the combined SDSS northern and southern sky coverage. We find that the total gas mass derived from the standard $\beta$-model corresponds to a baryon fraction of $\rho_{\mathrm{gas}}^{\mathrm{fil}} / \bar{\rho}_b = 0.127^{+0.019}_{-0.021}$, while the generalized $\beta$-model yields a consistent result of $\rho_{\mathrm{gas}}^{\mathrm{fil}} / \bar{\rho}_b = 0.118^{+0.021}_{-0.021}$.
The close agreement between the two models suggests that our estimates of the filamentary baryon content are robust with respect to assumptions about the gas distribution profile.

Moreover, when we extend the analysis to include a broader filament sample with lengths outside the $[30, 100]\,\mathrm{cMpc}$ range and the same redshift interval ($0.2 < z < 0.6$), comprising $N = 71,\!954$ filaments, the inferred gas fraction increases to $\rho_{\mathrm{gas}}^{\mathrm{fil}} / \bar{\rho}_b = 0.232^{+0.036}_{-0.035}$. This implies that up to 23\% of the cosmic baryon budget may reside in filamentary structures, further reinforcing the significance of the WHIM as a reservoir of missing baryons in the cosmic web.

\section{Discussion}
\label{sec:discussion}
A previous study conducted by Tanimura et al.~(2020)~\cite{Tanimura2020A&A...637A..41T} used the \textit{Planck} PR2 $y$-map and the CMB lensing convergence map to stack 24,544 filaments identified in the northern sky of SDSS DR12 (LOWZ-CMASS), adopting physical coordinates in their stacking procedure. Their best-fit results, based on a standard $\beta$-model, were $\delta = 19.0^{+27.3}_{-12.1}$, $r_c = 1.5^{+1.8}_{-0.7}\,\mathrm{Mpc}$, and $T_e = (1.4^{+0.4}_{-0.4}) \times 10^6\,\mathrm{K}$, are broadly consistent with ours, as both studies are based on similar datasets and adopt the same $\beta$-model framework.

Notably, our standard $\beta$-model yields a lower central overdensity and a larger core radius compared to Tanimura et al.~(2020)~\cite{Tanimura2020A&A...637A..41T}. This discrepancy likely arises from the intrinsic degeneracy in the standard $\beta$-model, where the fixed slope parameters $(\alpha, \beta) = (2, 2/3)$ allow a trade-off between central density and spatial extent to fit the observed profiles. The generalized $\beta$-model, in contrast, relaxes the slope constraints, enabling better characterization of the profile shape. It yields higher central overdensities and smaller core radii, more consistent with expectations for thermally and gravitationally confined filamentary gas. This highlights the value of flexible parameterizations when jointly modeling tSZ and lensing profiles.

Several other studies have investigated the baryon content in filaments using pair-based stacking techniques. Using the tSZ effect, Tanimura et al.~(2019a)~\cite{Tanimura2019aMNRAS.483..223T} stacked $\sim260{,}000$ SDSS-LOWZ galaxy pairs at $z \sim 0.3$ with transverse separations of $6$–$10\,h^{-1}\,\mathrm{Mpc}$, estimating $\delta \sim 3.2$. De Graaff et al.~(2019)~\cite{deGraff2019A&A...624A..48D} extended this method to $\sim10^6$ SDSS-CMASS pairs at $z \sim 0.55$ and found $\delta \sim 5.5$. In contrast, Epps \& Hudson (2017)~\cite{Epps2017MNRAS.468.2605E} analyzed the weak-lensing signal between $\sim23{,}000$ SDSS/BOSS pairs using the Canada-France-Hawaii telescope lensing survey(CFHTLenS) data, deriving an average filament mass of $(1.6 \pm 0.3) \times 10^{13}\,M_\odot$ in a $7.1 \times 2.5\,h^{-1}\,\mathrm{Mpc}$ region, corresponding to $\delta \sim 4$ under the assumption of uniform density.

In simulations, T.Tuominen et al.~(2021)~\cite{Tuominen2021A&A...646A.156T} used the Bisous method in the EAGLE simulation and found central radii of $r \sim 1\,\mathrm{Mpc}$, with filamentary gas in the hot WHIM temperature range ($>10^{5.5}\,\mathrm{K}$) and overdensity $\delta \sim 1$–$50$.
W.Zhu et al.~(2021)~\cite{zhu2021ApJ...920....2Z} studied cosmic filaments using a cosmological hydrodynamic simulation with adaptive mesh refinement (AMR). For the filament group with a radius of $4\,h^{-1}\mathrm{Mpc}$, the central overdensity lies between $4$ and $5$, with a core radius of approximately $3.2\,h^{-1}\mathrm{Mpc}$ and a temperature of $4 \times 10^6\,\mathrm{K}$.
Similarly, for the $3\,h^{-1}\mathrm{Mpc}$ group, the central overdensity is also in the range of $4$–$5$, with a core radius of about $2.4\,h^{-1}\mathrm{Mpc}$ and a temperature of $2 \times 10^6\,\mathrm{K}$.
Galárraga-Espinosa et al.~(2022)~\cite{Galarraga-Espinosa2022A&A...661A.115G} studied long DisPerSE filaments ($l > 20\,\mathrm{Mpc}$) in the IllustrisTNG simulation, reporting a gas overdensity of $1+\delta = 19.34 \pm 1.28$, a core radius of $r_0 = 0.51 \pm 0.11\,\mathrm{Mpc}$, and a temperature range of $10^5$–$10^{6.5}\,\mathrm{K}$.
In the recent work by Qi-Rui Yang et al.~(2025)~\cite{yang2025arXiv250702476Y}, their thickest filaments (identified with DisPerSE in various runs of the TNG simulations), with radii of $2$–$3.2\,\mathrm{cMpc}$, show $1+\delta \sim 8$ and temperatures in the range $10^{5.5}$–$10^{6}\,\mathrm{K}$.
Although these filament samples in simulation are generally shorter and thinner than those in our work, the results remain consistent.

Among these studies, Tanimura et al.~(2020)~\cite{Tanimura2020A&A...637A..41T} provided an explicit estimate of the filamentary contribution to the cosmic baryon budget. They reported a baryon fraction of $0.080^{+0.116}_{-0.051} \times \Omega_b$ for filaments in the $30$–$100\,\mathrm{Mpc}$ range, and suggested a possible increase to $\sim0.141 \times \Omega_b$ when extrapolating to include shorter and longer filaments under the assumption of universal gas density. In contrast, our analysis yields a higher fraction of $0.127^{+0.019}_{-0.021} \times \Omega_b$ within the $30$–$100\,\mathrm{cMpc}$ length range, which increases to $\sim 0.232 \times \Omega_b$ when extended to filaments of all lengths in our sample.

These improvements stem from several methodological advancements in our analysis. First, we utilize the updated \textit{Planck} PR4 $y$-map, which benefits from reduced noise and improved calibration relative to PR2. Second, our filament catalog covers both the northern and southern skies, in contrast to previous studies confined to the north, thereby enhancing the statistical completeness of the sample. Third, our stacking procedure traces the exact HEALPix pixel positions along each filament spine in comoving coordinates, and incorporates continuity-preserving treatments at junctions between filament segments, avoiding artifacts introduced by grid interpolation. Finally, we apply a geometric projection bias correction based on filament inclination, mitigating signal loss from LOS orientations. This correction helps recover the underestimated signals. These enhancements collectively contribute to the higher detection significance, as well as the increased estimates of both the central electron temperature ($T_e$) and the baryon fraction in our analysis. In particular, the inclusion of geometric bias corrections during model fitting may partially explain the elevated $T_e$ values compared to Tanimura et al.~(2020)~\cite{Tanimura2020A&A...637A..41T}.

The overall consistency of our baryon fraction estimate ($0.127^{+0.019}_{-0.021} \times \Omega_b$) with that of de Graaff et al.~(2019)~\cite{deGraff2019A&A...624A..48D}, who reported $(0.11 \pm 0.07) \times \Omega_b$ based on an independent pair-based tSZ analysis, further reinforces the role of the WHIM in filaments as a major reservoir of the missing baryons in the low-redshift Universe. Furthermore, by including filaments outside the $30$–$100\,\mathrm{cMpc}$ range, our estimated baryon fraction increases to ($0.232^{+0.036}_{-0.035} \times \Omega_b$), which agrees well with the value of $(28 \pm 12)\% \times \Omega_b$ inferred by de Graaff et al.~(2019)~\cite{deGraff2019A&A...624A..48D} when averaging over longer galaxy-pair baselines. Similarly, Tuominen et al. (2021)~\cite{Tuominen2021A&A...646A.156T} reported that the detected filaments in the EAGLE simulation account for approximately $23$–$25\%$ of the total cosmic baryon budget.

\section{Conclusion}
\label{sec:conclusion}

In this work, we investigated the distribution of the missing baryons in the cosmic web by analyzing the tSZ and CMB lensing signals associated with large-scale filaments. We identified filaments in the SDSS LOWZ-CMASS sample using the DisPerSE algorithm, and constructed a filament catalog covering both the northern and southern Galactic sky regions, a redshift range of $ 0.2 \sim 0.6 $, and a length range of $30 \sim 100\,\mathrm{cMpc}$. To isolate the diffuse filamentary signal, we applied comprehensive masking to remove resolved galaxy clusters and foreground contamination, including masking out to $3\times R_{500}$ for known halos from multiple cluster catalogs. By further excluding filaments with extreme inclination angles to avoid significant projection contamination, a total of $\sim 30,700$ filaments remain for the final profile modeling and parameter fitting.

We stacked the \textit{Planck} PR4 Compton-$y$ and CMB lensing convergence ($\kappa$) maps in comoving coordinates around the filament spines, yielding high signal-to-noise average radial profiles of both observables ($7.82\sigma$ for the $y$ profile and $7.78\sigma$ for the $\kappa$ profile). By jointly fitting the stacked $y$ and $\kappa$ profiles with both a standard and a generalized $\beta$-model, we constrained the central gas properties of filaments. The standard model yields 
$\delta = 4.18^{+2.01}_{-1.06}$, 
$T_e = (2.74^{+0.65}_{-0.53}) \times 10^6\,\mathrm{K}$, 
and $r_c = 5.18^{+1.20}_{-1.39}\,\mathrm{cMpc}$,
while the generalized model returns a higher central density 
$\delta = 7.49^{+8.88}_{-2.87}$ 
and a smaller core radius 
$r_c = 3.28^{+2.16}_{-1.72}\,\mathrm{cMpc}$,
with a slightly elevated temperature of 
$T_e = (3.06^{+0.78}_{-0.60}) \times 10^6\,\mathrm{K}$.

For filaments in the $30$–$100\,\mathrm{cMpc}$ length range, our analysis yields consistent baryon fractions from both models: $0.127^{+0.019}_{-0.021} \times \Omega_b$ from the standard $\beta$-model and $0.118^{+0.021}_{-0.021} \times \Omega_b$ from the generalized $\beta$-model. This agreement indicates that our gas fraction estimates are robust against assumptions about the underlying profile shape. When extending the analysis to include filaments outside this length range, the inferred baryon fraction rises to $0.232^{+0.036}_{-0.035}$. These results strengthen the interpretation that the WHIM in filaments hosts a substantial fraction of the missing baryons in the low-redshift Universe.

Our work improves upon earlier studies by employing the latest \textit{Planck} $y$-map data (PR4), incorporating both northern and southern sky coverage, and refining the stacking methodology in comoving coordinates. The joint modeling of $y$ and $\kappa$ profiles enables tighter constraints on gas overdensity and temperature, with the generalized $\beta$-model alleviating degeneracies present in the standard formulation.

Future studies using higher-resolution tSZ and lensing datasets, such as from Simons Observatory~\cite{AdeAguirre2019JCAP...02..056A} or CMB-S4~\cite{Abazajian2016arXiv161002743A}, will enable filament-by-filament analyses and provide even tighter constraints on the physical state of diffuse baryons in the cosmic web.

\acknowledgments
We thank Peng Wang and Yanchuan Cai for helpful discussions. We thank the anonymous referee for useful suggestions that significantly improve the quality of paper. This work is supported by the National SKA Program of China (2025SKA0150104) and the National Natural Science Foundation of China (NFSC) through grant 12203107, 11733010 and 12173102.

\appendix

\section{Impact of the FoG effect}
\label{app:fog}
\begin{figure}[t!]
    \centering
    \includegraphics[width=0.8\textwidth]{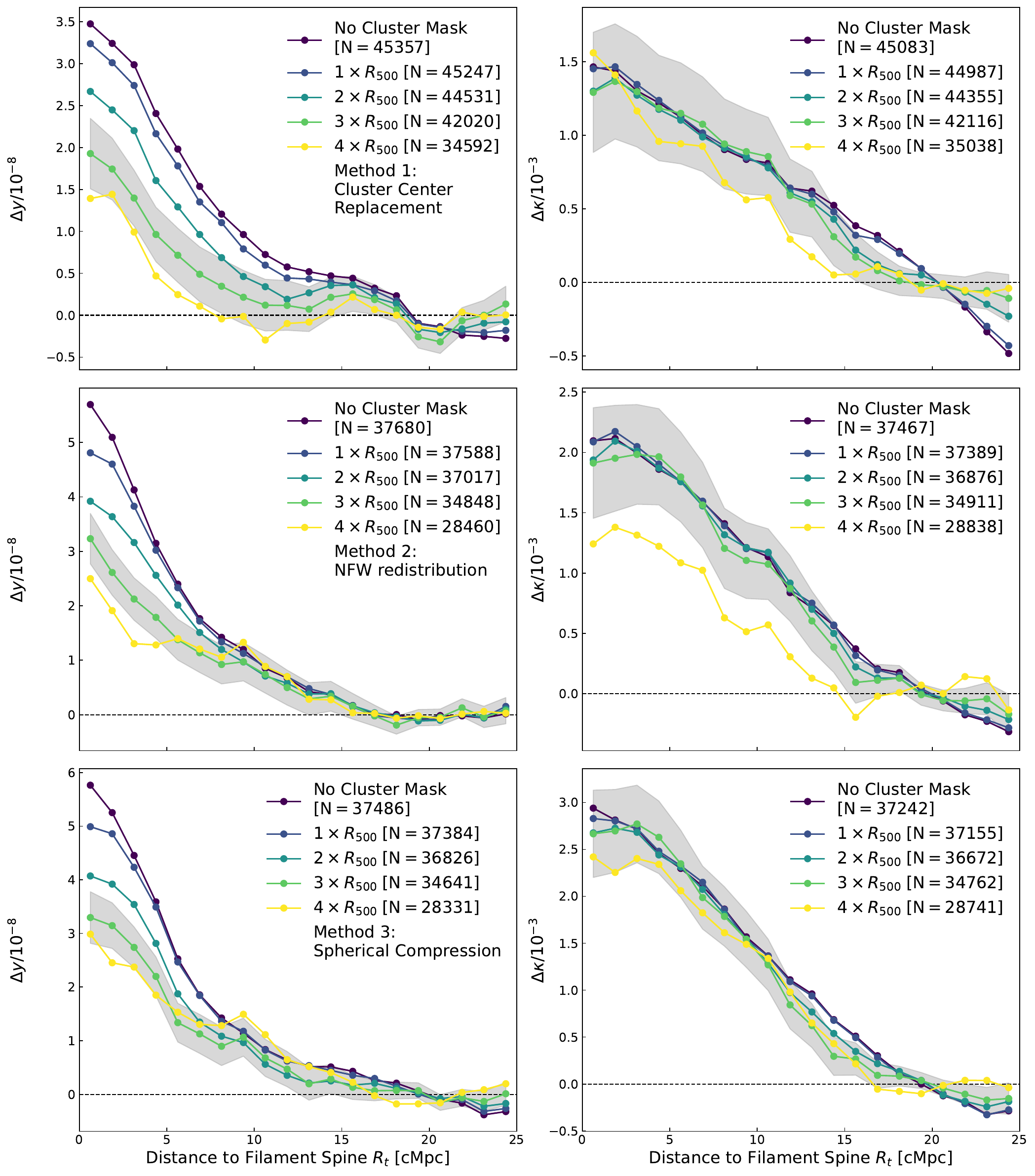}
    \caption{
Stacked radial profiles of Compton-$y$ (\textbf{left column}) and CMB-lensing convergence $\kappa$ (\textbf{right column}) obtained after applying three different FoG correction strategies.  
\textbf{Top}: Cluster-center replacement, where galaxies inside the FoG cylinder are removed and substituted with a single representative object at the cluster center.  
\textbf{Middle}: NFW redistribution, where galaxies in the FoG cylinder are re-sampled within a spherical region according to an NFW radial probability distribution.
\textbf{Bottom}: Spherical compression, in which only the LOS elongated component of the FoG cylinder is collapsed onto a spherical surface of radius $3R_{\rm vir}$.  
}
    \label{fig:fog_y_kappa_profiles}
\end{figure}

\jz{To assess the impact of the FoG effect on our filament reconstruction, we performed a series of dedicated tests. For each cluster in our sample, we defined a cylindrical region (FoG cylinder) with radius $R_{\rm cyl} = 3R_{\rm vir}$ and half height $dz = 3(\sigma_v/c)(1+z)$, where the velocity dispersion is estimated as $\sigma_v=\sqrt{GM_{\rm vir}/(5R_{\rm vir})}$. All galaxies located within this cylinder were subjected to one of the following three FoG correction strategies:}

\jz{ \textbf{Method 1: Cluster center replacement.}  
Following the prescription of Malavasi et al. (2020)~\cite{Malavasi2020A&A...642A..19M}, all galaxies inside the cylinder were removed. The average DTFE estimated density of the removed galaxies, $\rho_{\rm removed}$, was computed, and a single mock galaxy was placed at the cluster center (the centroid of the cylinder). This mock object was assigned the density $\rho_{\rm removed}$, preserving the local mass contribution. The DTFE density for all remaining galaxies is kept unchanged. A new tessellation is constructed using the modified catalog, and DisPerSE is run on this updated density field. This procedure eliminates the artificial LOS elongation while preserving the total mass contribution of the region.}

\jz{\textbf{Method 2: NFW redistribution.}  
To provide a more physically realistic correction, we considered a redistribution scheme in which galaxies inside the FoG cylinder were re-positioned within the same spherical region of radius $3R_{\rm vir}$, but their radial distances were drawn from an NFW probability distribution. This yields a galaxy configuration consistent with the expected dark matter dominated halo density profile, thereby mimicking a more realistic post FoG-correction configuration.} 

\jz{\textbf{Method 3: Spherical compression.}  
Actually not all galaxies within the FoG cylinder belong to clusters in real space. By placing all of them into the cluster-centered $3R_{\rm vir}$ sphere, we artificially overestimate the density of the cluster region, which might cause the poorer convergence seen specifically in the case of Method 1. To address this, 
we apply a non-physical compression scheme which decreases the level of galaxy position compression in the FoG correction, as compared to Method 1 and 2. 
Galaxies located within the FoG cylinder but lying outside a sphere of radius $3R_{\rm vir}$ were radially projected onto the surface of this sphere, while galaxies already inside the sphere were left unchanged. In this way, the intrinsically spherical inner halo structure is preserved, whereas the spurious LOS-extended component generated by the FoG effect is removed. This method provides an intentionally conservative deformation designed to test whether such elongated features could bias the subsequent filament extraction.}

\jz{For each of the three approaches, the modified galaxy positions and DTFE densities were supplied to DisPerSE to re-extract the filament network. We then repeated the stacking analysis of the Compton-$y$ and CMB-lensing $\kappa$ profiles using the re-generated filament catalogs.}

\jz{Figure~\ref{fig:fog_y_kappa_profiles} presents the stacked Compton-$y$ and $\kappa$ profiles obtained after applying the three FoG–mitigation strategies, evaluated under different cluster–masking radii. In cases where a larger mask of $4\times R_{500}$ is applied, the profiles exhibit reduced signal amplitude and, in some instances, poorer convergence. This is likely due to the fact that the three FoG correction schemes, while useful as robustness tests, do not necessarily reconstruct the exact physical positions of galaxies within clusters, and therefore may shift the reconstructed filament spines slightly.}

\jz{However, for most cases the profiles derived from Methods~2 and~3 remain broadly consistent with our fiducial results obtained from the original galaxy catalog with a standard $3\times R_{500}$ cluster mask. In addition, our stacking procedure already removes cluster-dominated regions through masking, and the fitting is further restricted to angles that exclude the LOS direction, where FoG induced artificial structures would appear. These two layers of mitigation effectively suppress spurious LOS-aligned features.}

\jz{Taken together, these tests indicate that although the geometric modifications applied in the FoG-corrected catalogs may alter the reconstructed skeleton locally, the overall stacked $y$ and $\kappa$ profiles remain stable. We therefore conclude that the FoG effect does not significantly bias either the filament identification or the resulting stacked measurements. Nevertheless, more physically motivated or higher-fidelity FoG correction techniques could be explored in future work to further refine the reconstruction, particularly in dense cluster environments where peculiar velocities are large.}

\section{Robustness to background subtraction range}
\label{app:bg}
\begin{figure}[t!]
    \centering
    \begin{minipage}[t]{0.49\textwidth}
        \centering
        \includegraphics[width=\textwidth]{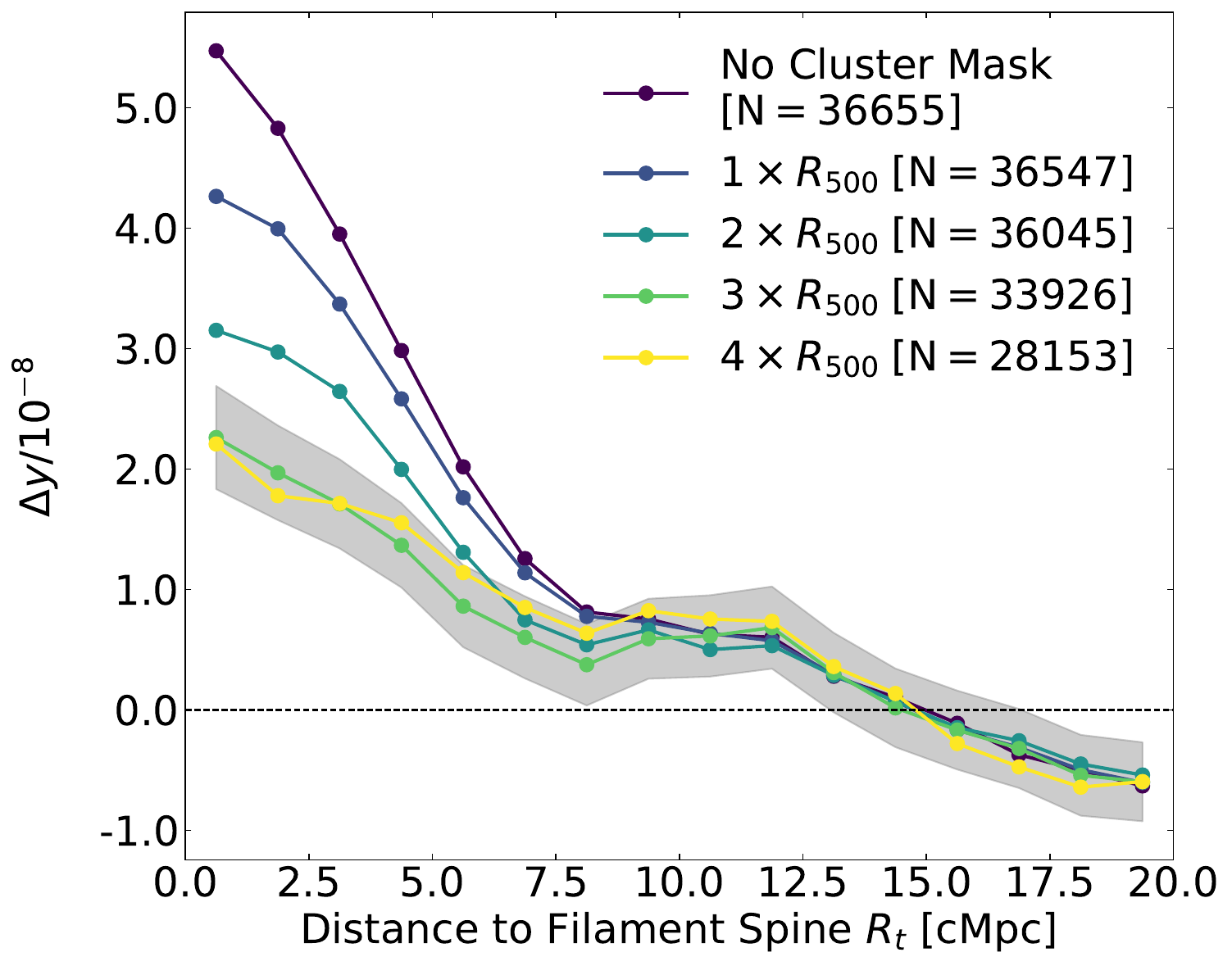}
    \end{minipage}\hfill
    \begin{minipage}[t]{0.49\textwidth}
        \centering
        \includegraphics[width=\textwidth]{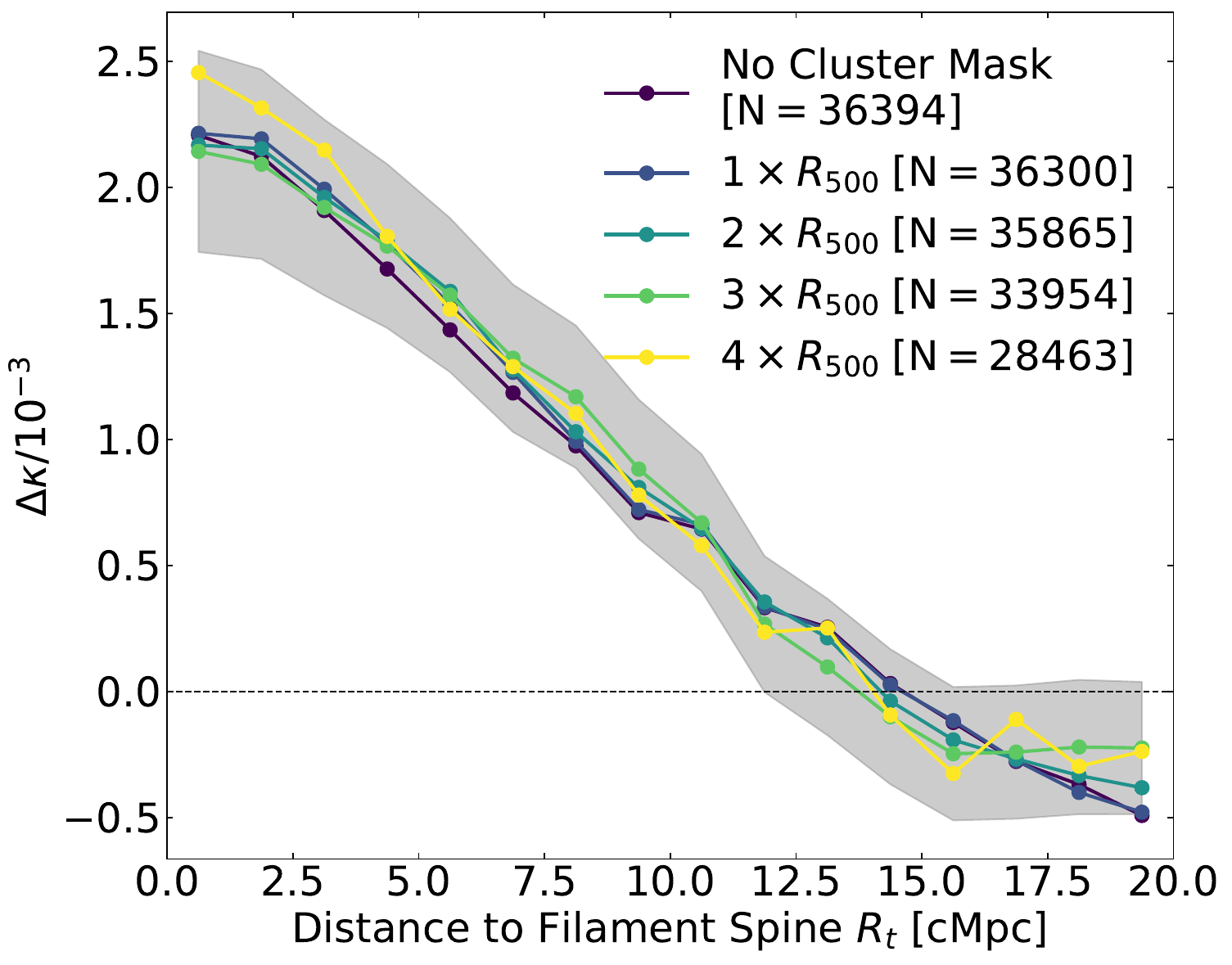}
    \end{minipage}
    \caption{
Stacked radial profiles using a background subtraction range of $10$--$20\,\mathrm{cMpc}$. 
\textbf{Left}: Compton-$y$ profiles.
\textbf{Right}: CMB lensing convergence ($\kappa$) profiles.
}
    \label{fig:stack_bg1020}
\end{figure}
\begin{figure}[t!]
    \centering
    \begin{minipage}[t]{0.48\textwidth}
        \centering
        \includegraphics[width=\textwidth]{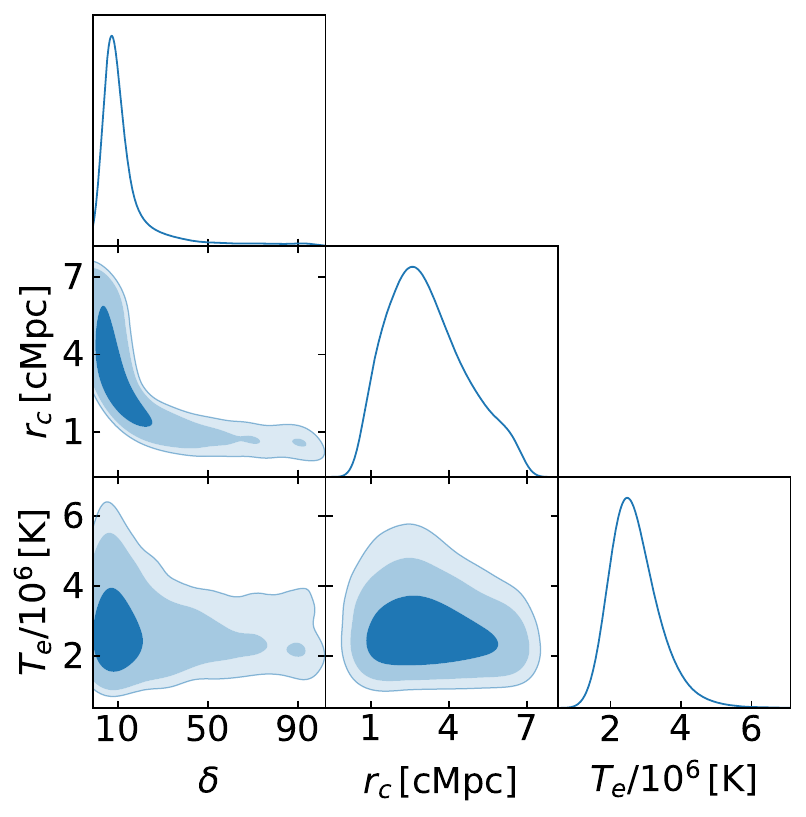}
    \end{minipage}\hfill
    \begin{minipage}[t]{0.48\textwidth}
        \centering
        \includegraphics[width=\textwidth]{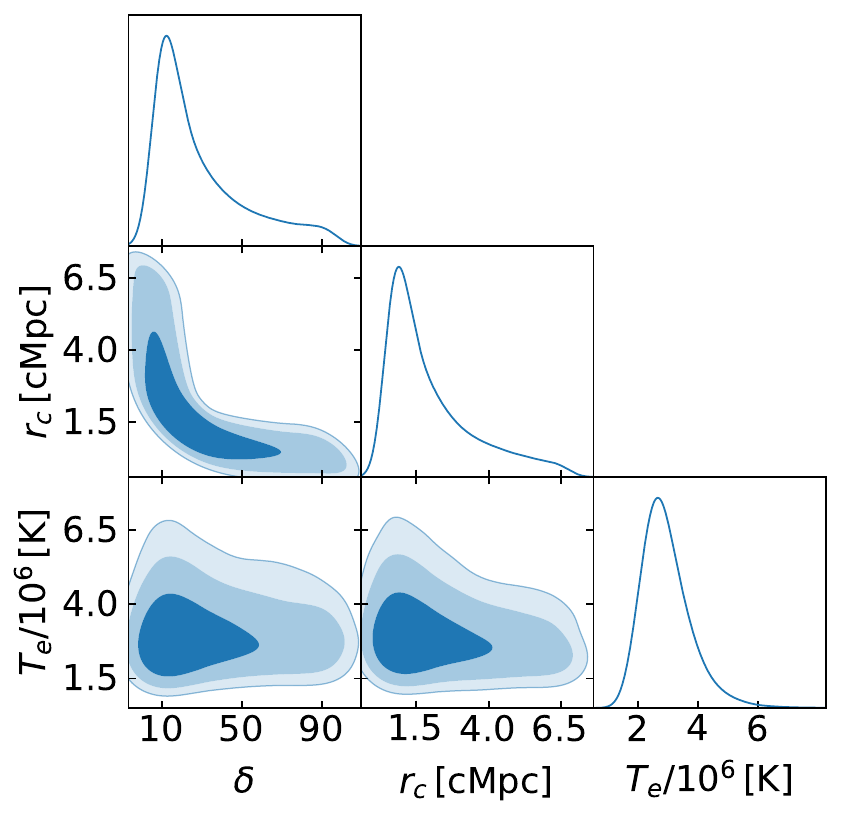}
    \end{minipage}
    \caption{
Corner plots of the posterior distributions for the model parameters obtained from the MCMC fitting of the stacked $y$ and $\kappa$ profiles, using the $10$–$20~\mathrm{cMpc}$ background subtraction range.
\textbf{Left}: Standard $\beta$-model.
\textbf{Right}: Generalized $\beta$-model.
The diagonal panels display the marginalized one-dimensional posterior distributions for the central overdensity $\delta$, core radius $r_c$, and electron temperature $T_e$. The off-diagonal panels show the joint two-dimensional posteriors, with contours indicating the $1\sigma$, $2\sigma$, and $3\sigma$ credible regions.
}
    \label{fig:corner_bg1020}
\end{figure}

\begin{figure}[t!]
    \centering
    \begin{minipage}[t]{0.49\textwidth}
        \centering
        \includegraphics[width=\textwidth]{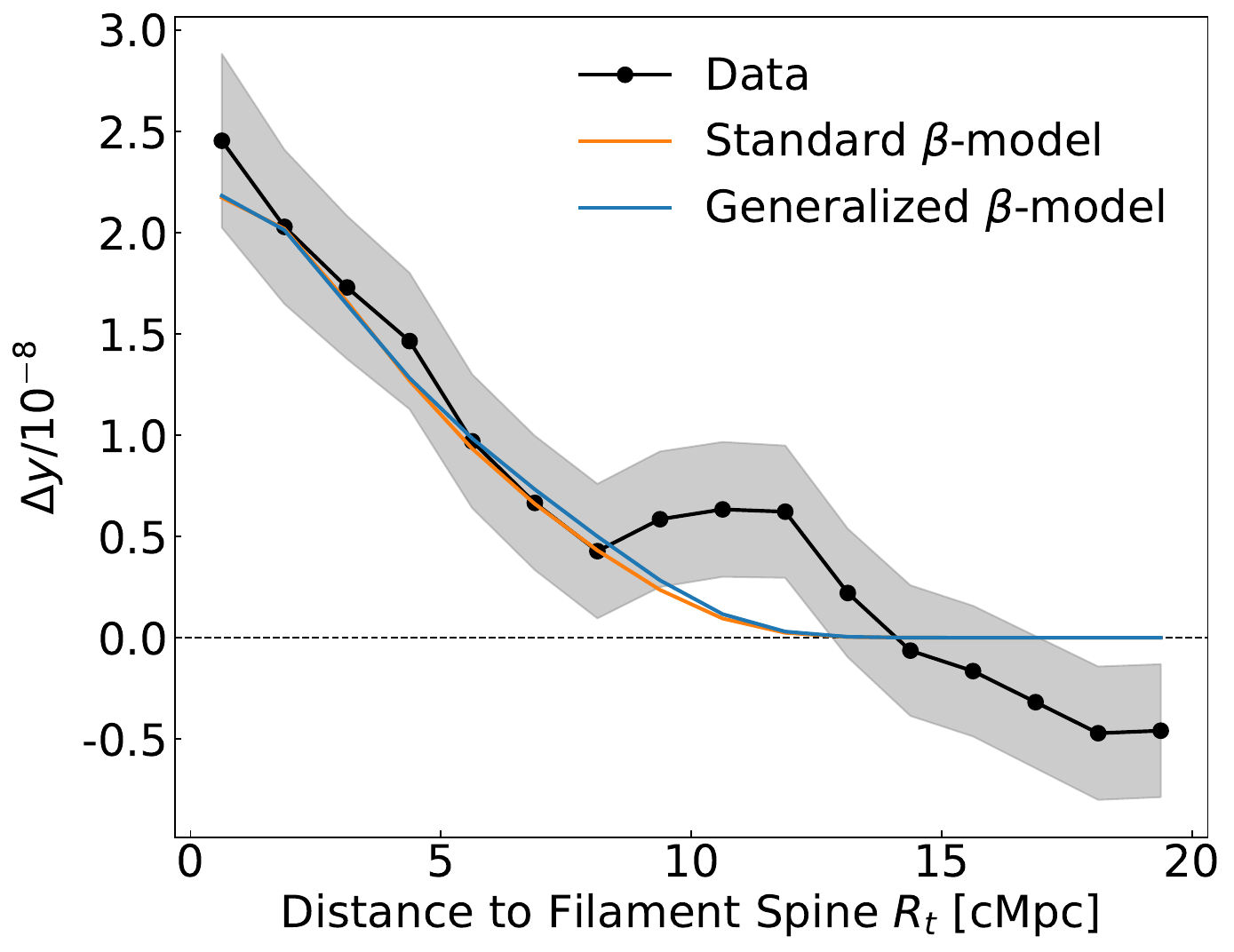}
    \end{minipage}\hfill
    \begin{minipage}[t]{0.49\textwidth}
        \centering
        \includegraphics[width=\textwidth]{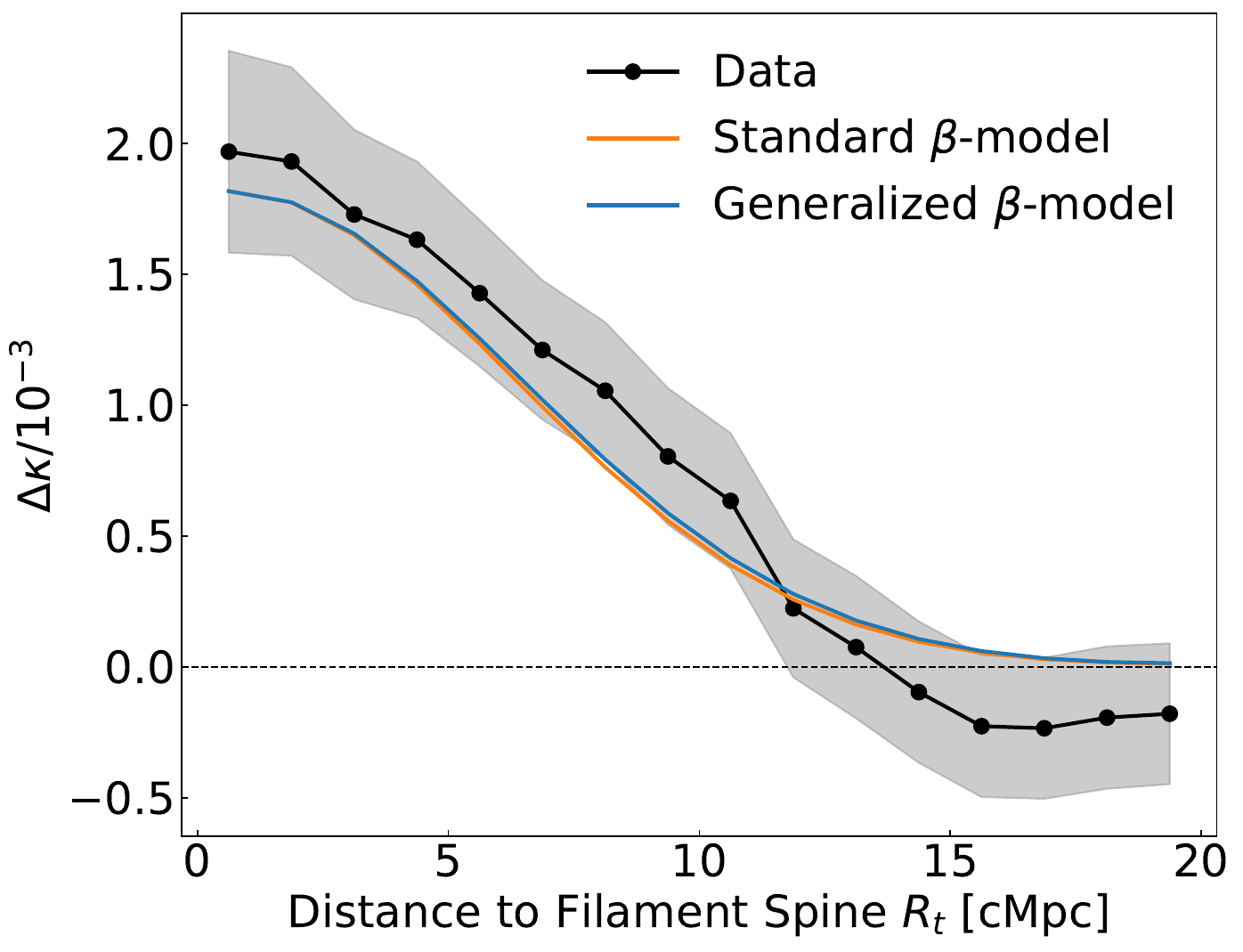}
    \end{minipage}
    \caption{
Average radial profiles of filaments obtained using the $10$–$20~\mathrm{cMpc}$ background subtraction range. \textbf{Left}: Compton-$y$ profiles. \textbf{Right}: CMB lensing convergence ($\kappa$) profiles. The profiles are fitted with the standard $\beta$-model (orange) and the generalized $\beta$-model (blue).
}
    \label{fig:fit_bg1020}
\end{figure}

To assess the robustness of our stacking analysis with respect to the choice of background region, we repeated the procedure using a background subtraction range of $10$--$20\,\mathrm{cMpc}$ from the filament spine, instead of the fiducial $15$--$25\,\mathrm{cMpc}$ range adopted in the main analysis. All other aspects of the analysis, including masking, coordinate definitions, and model fitting procedures, were kept identical.

Figure~\ref{fig:stack_bg1020} presents the stacked $y$ and $\kappa$ profiles obtained using this alternative background range. While the profiles remain consistent with the fiducial results within uncertainties, their overall amplitudes are slightly lower. This difference arises because the $10$--$20\,\mathrm{cMpc}$ region contains a higher average signal than the $15$--$25\,\mathrm{cMpc}$ range, resulting in stronger background subtraction and a reduced net signal.

To ensure consistency, we maintain the same model fitting procedure and pipeline as adopted in the fiducial background subtraction case ($15$--$25\,\mathrm{cMpc}$), including the MCMC setup, parameter priors, and likelihood definitions. For both the standard and generalized $\beta$-models, we fix the integration limit to $R_{\mathrm{max}} = 10~\mathrm{cMpc}$ to match the reduced background subtraction range. This allows for a direct comparison of the best-fit results and the corresponding baryon fractions with those obtained under the fiducial setup.

The posterior distributions of the model parameters, obtained from the MCMC fitting, are shown in Figure~\ref{fig:corner_bg1020}. The corresponding best-fit profiles are overlaid in Figure~\ref{fig:fit_bg1020} for both the $y$ and $\kappa$ signals. The best-fit parameters derived from this test are summarized in Table~\ref{tab:fit_bg1020}. The corresponding baryon fractions—computed by integrating the fitted profiles up to $R_{\max} = 10\,\mathrm{cMpc}$ to match the fitting range—are $\rho_{\mathrm{gas}}^{\mathrm{fil}} / \bar{\rho}_b = 0.086^{+0.016}_{-0.017}$ for the standard $\beta$-model and $\rho_{\mathrm{gas}}^{\mathrm{fil}} / \bar{\rho}_b = 0.084^{+0.016}_{-0.017}$ for the generalized $\beta$-model. These values are slightly lower than those obtained using the fiducial background, as expected. This is primarily due to the subtraction of a higher background level and the adoption of a smaller cutoff radius. As a result, the recovered baryon fractions are lower but more consistent with those reported in Tanimura et al. (2020)~\cite{Tanimura2020A&A...637A..41T}.

\begin{table}[t!]
\renewcommand{\arraystretch}{1.5} 
\centering
\begin{tabular}{
    >{\columncolor{{C0}!20}}l |
    >{\columncolor{{C0}!10}}c |
    >{\columncolor{{C0}!10}}c }
\rowcolor{{C0}!25}
\hline
Parameter & Standard $\beta$-model & Generalized $\beta$-model \\
\hline
$\delta$ & $8.04^{+10.74}_{-3.37}$ & $19.55^{+32.88}_{-11.51}$ \\
$r_{c}~[\mathrm{cMpc}]$ & $2.98^{+1.84}_{-1.37}$ & $1.40^{+2.11}_{-0.80}$ \\
$T_e~[10^6~\mathrm{K}]$ & $2.62^{+0.78}_{-0.60}$ & $2.83^{+0.89}_{-0.67}$ \\
\hline
\end{tabular}
\caption{Best-fit parameters for the standard and generalized $\beta$-models from the joint MCMC fit to the $y$ and $\kappa$ profiles using a background subtraction range of $10$--$20\,\mathrm{cMpc}$. Uncertainties represent the 16th and 84th percentiles of the marginalized posterior distributions. \label{tab:fit_bg1020}}
\end{table}

\section{Redshift, length, and environment dependence of the stacked signal}
\label{app:subsamples}
\begin{figure}[t!]
    \centering
    \includegraphics[width=0.8\textwidth]{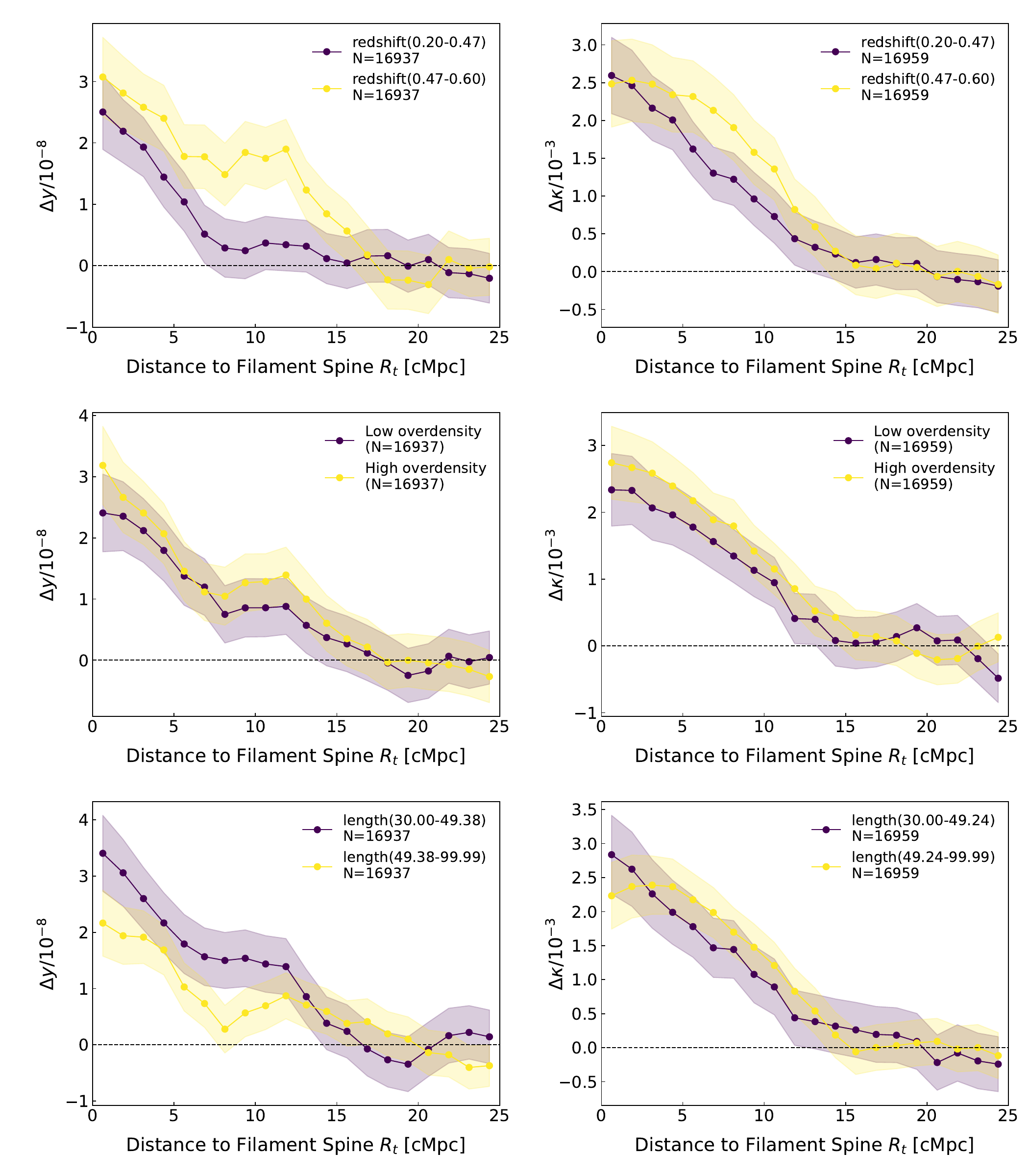}
    \caption{
    Stacked radial profiles of Compton-$y$ (\textbf{left column}) and CMB-lensing convergence $\kappa$ (\textbf{right column}) for different filament subsamples.  
    \textbf{Top}: Filaments split into two redshift bins, $0.2 < z < 0.47$ and $0.47 < z < 0.6$.  
    \textbf{Middle}: Filaments in low-density and high-density environments.  
    \textbf{Bottom}: Filaments divided by length.
    }
    \label{fig:subsample_yk_profiles}
\end{figure}

To explore possible redshift evolution and environmental dependence of the WHIM properties, we repeat the stacking analysis for three filament subsamples defined by redshift, length, and environmental overdensity. 
The subsample definitions are as follows:

\begin{itemize}
    \item \textbf{Redshift bins:} The filament sample is divided into two subsamples of equal size by the median redshift: a low-redshift bin ($0.2 < z < 0.47$) and a high-redshift bin ($0.47 < z < 0.6$).
    
    \item \textbf{Environmental overdensity bins:} We divide the sample according to the local matter overdensity at filament endpoints, estimated using the DTFE implementation in DisPerSE. Two equal-number subsamples are defined based on the overdensity quantiles, corresponding to low-density and high-density environments.
    
    \item \textbf{Length bins:} The filaments are further divided into two equal-sized subsamples based on their median physical length of approximately $50\,\mathrm{cMpc}$.
\end{itemize}

For each subsample, we extract the radial Compton-$y$ and CMB-lensing convergence ($\kappa$) profiles using the same stacking procedure as described in the main text. The resulting profiles are shown in Fig.~\ref{fig:subsample_yk_profiles}.

\paragraph{Redshift dependence.}
We find a marginally enhanced $y$-signal in the high-redshift bin ($0.47 < z < 0.6$), which appears to drive the bump observed around $10\,\mathrm{cMpc}$. This feature may arise from the contribution of neighbouring filaments that enter the stacking aperture at this projected separation, although confirming this interpretation will require further investigation.


\paragraph{Environmental dependence.}
Filaments residing in high-density environments exhibit a modest enhancement in both the tSZ and CMB-lensing signals compared with those in low-density environments. The overall profiles, however, remain consistent within the bootstrap uncertainties.

\jz{\paragraph{Length dependence.}
We observe that shorter filaments exhibit a stronger tSZ signal. This trend may reflect the fact that very long filaments tend to include merged or noisy segments, which dilute the averaged stacked profile.
Overall, the subsample analyses demonstrate that the stacked signals are robust to variations in filament redshift, length, and environmental overdensity. The qualitative trends observed across different subsamples are consistent with physical expectations of gas thermal properties and matter distribution within the cosmic web.}

\section{Geometric correction for filament inclination}
\label{app:geometry_bias}
\begin{figure}[t!]
    \centering
    \includegraphics[width=0.8\textwidth]{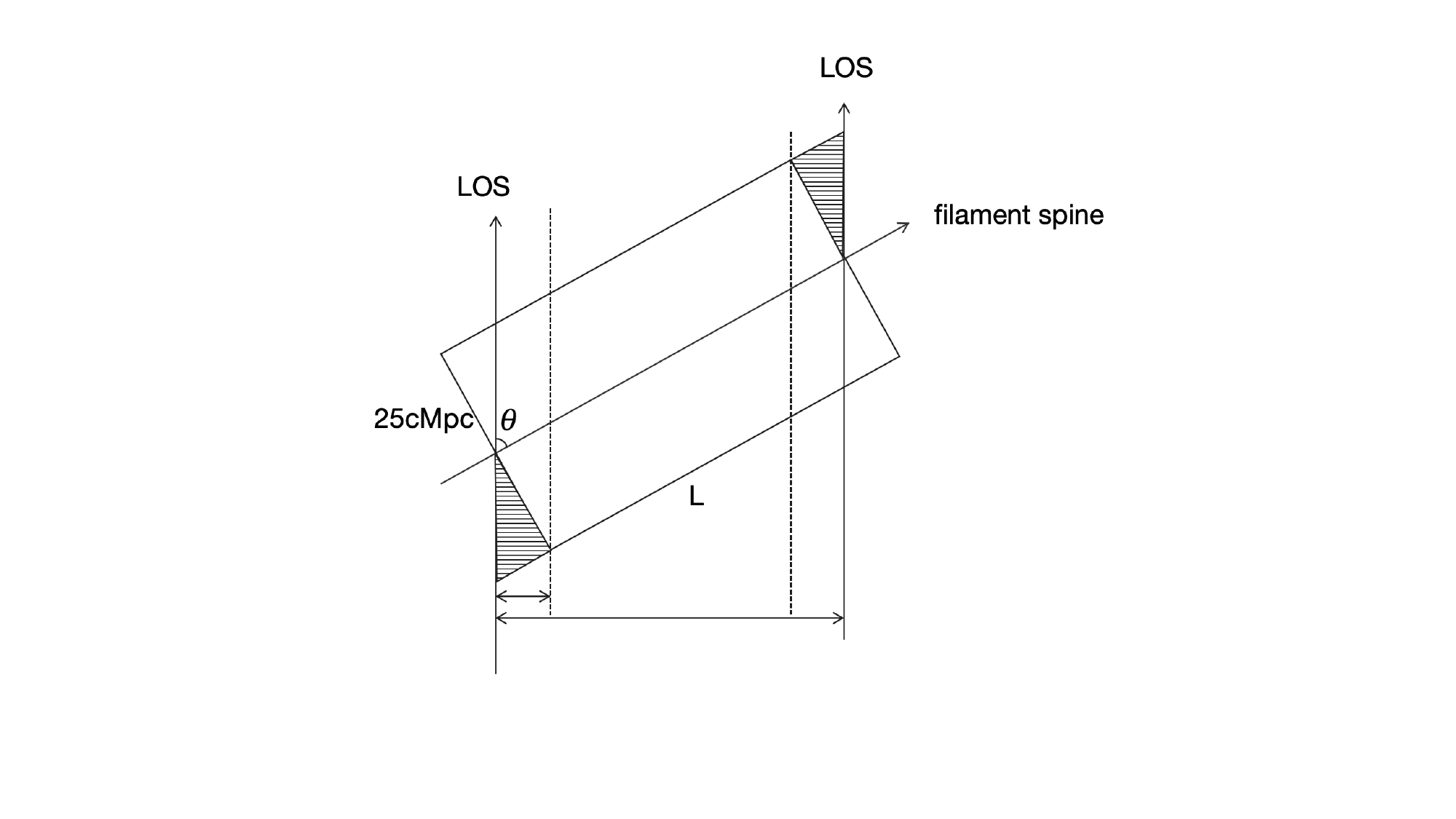}
    \caption{A schematic top-down view illustrating the geometric bias due to filament inclination. The shaded triangular regions represent projected areas outside the true filament that contribute to the observed signal.}
    \label{fig:bias_diagram}
\end{figure}

\begin{figure}[t!]
    \centering
    \begin{subfigure}[t]{0.49\textwidth}
        \centering
        \includegraphics[width=\textwidth]{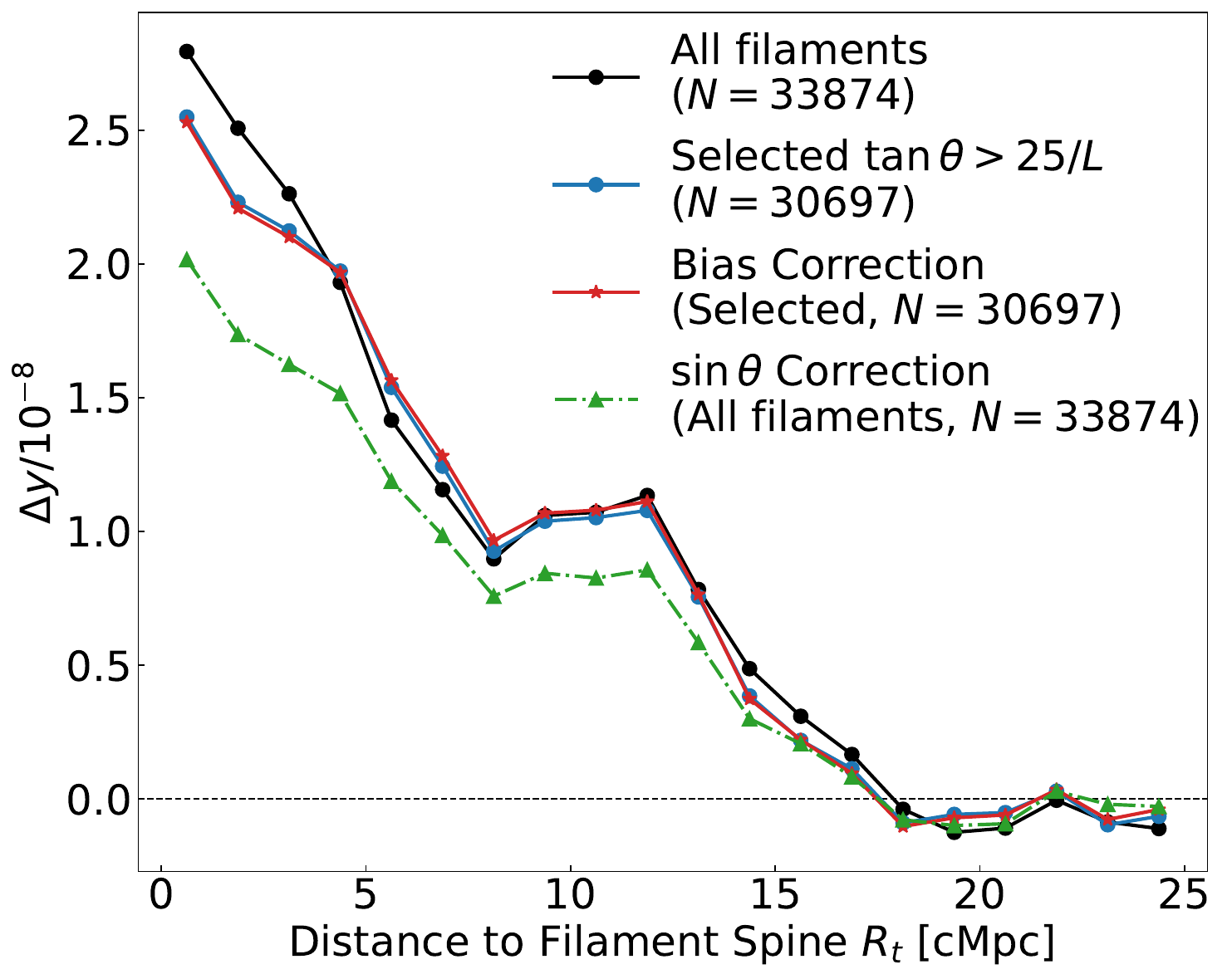}
        \caption{}
    \end{subfigure}
    \hfill
    \begin{subfigure}[t]{0.49\textwidth}
        \centering
        \includegraphics[width=\textwidth]{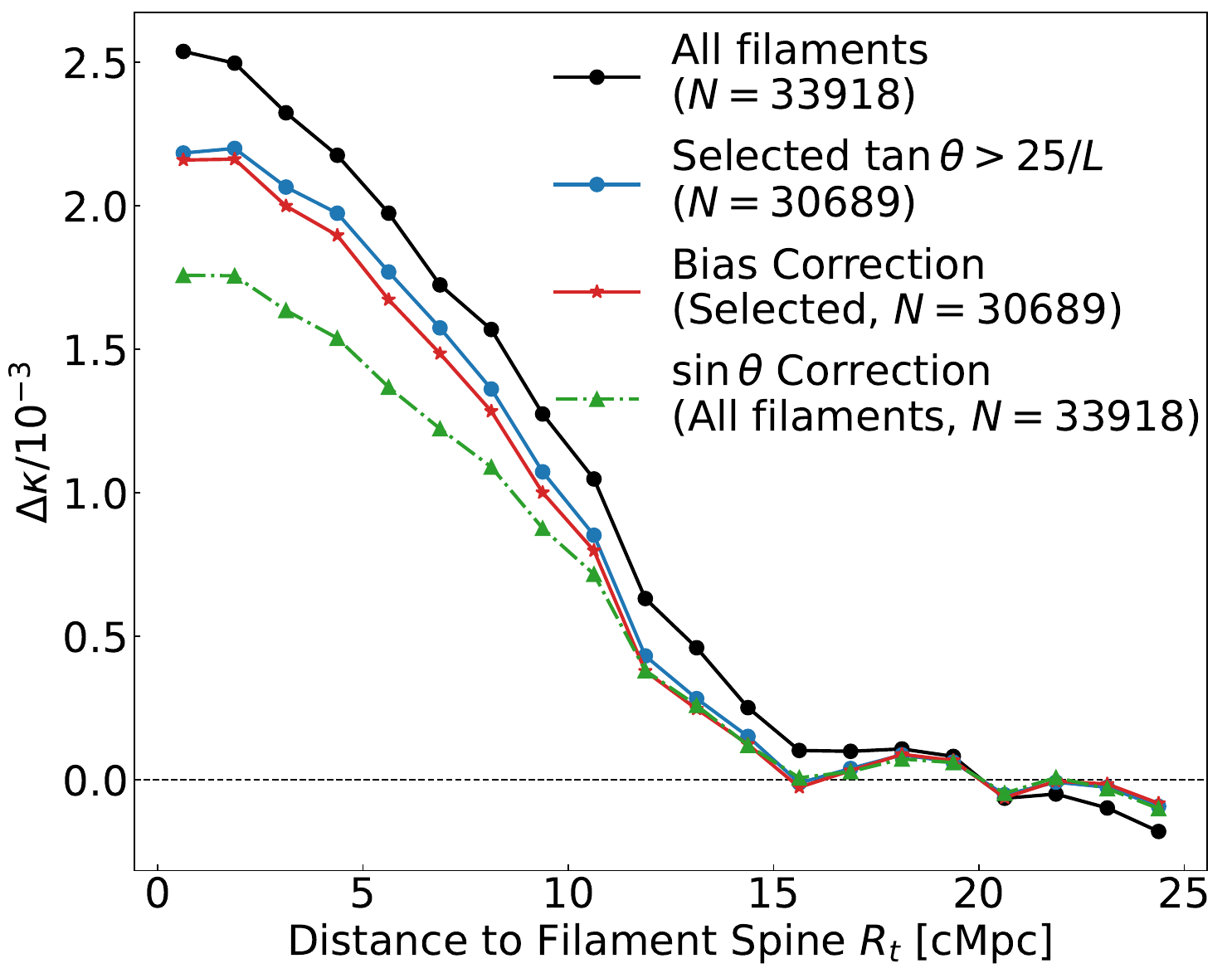}
        \caption{}
    \end{subfigure}
    \caption{Comparison of different methods for correcting geometric bias in the stacked radial Compton-$y$ profile \textbf{(a)} and lensing convergence $\kappa$ profile \textbf{(b)}. The black curve shows the original (uncorrected) signal. The blue curve excludes filaments with $\tan \theta < 25 / L$ to eliminate extreme projection contamination. The red curve applies the refined bias correction described in eq.~\eqref{eq:signal_correction} to the filtered filament subset. The green curve applies the $\sin \theta$-based correction adopted by Tanimura et al. (2020)~\cite{Tanimura2020A&A...637A..41T}.
}
    \label{fig:bias_correction}    
\end{figure}

In our modeling framework, we assume an idealized scenario in which each filament is modeled as an infinite, isothermal, cylindrically symmetric structure oriented perpendicular to the LOS. In reality, however, observed filaments exhibit a range of inclinations relative to the plane of the sky, introducing projection effects that must be corrected.

Tanimura et al. (2020)~\cite{Tanimura2020A&A...637A..41T} addressed filament inclination by explicitly incorporating the orientation angle into their model. In contrast, we implement a geometric correction directly on the observed signal. We define the inclination angle $\theta$ for each filament as the angle between the straight line connecting its two endpoints (filament spine) and LoS. This definition assumes the filament is globally straight and does not account for local curvature or segment-level deviations.

To transform the measured signal into the equivalent for a filament perpendicular to the LoS, we multiply the observed Compton-$y$ signal by $\sin \theta$ during model fitting. However, this correction alone is insufficient. Due to the finite filament length  $L$ (average $\sim50\,\mathrm{cMpc}$) and the fixed measurement region ($50\,\mathrm{cMpc}$), projection effects cause over-integration near filament ends. As illustrated in Figure~\ref{fig:bias_diagram}, the projected stacking cylinder includes two triangular regions at each end that lie outside the physical filament but still contribute to the observed signal.

We estimate the fractional signal contamination from these over-projected areas using geometric considerations. The bias fraction is given by:
\begin{equation}
\mathrm{bias} = \frac{2 \cdot (25 / \tan \theta) \cdot \sin \theta \cdot \frac{1}{4}}{L \cdot \sin \theta} = \frac{25 \cos \theta}{2L \sin \theta}\,,
\label{eq:bias_formula}
\end{equation}
where $L$ is the three-dimensional length of the filament, and $\theta$ is the inclination angle. \jz{This expression accounts for the fact that, within the projected filament extent $L\sin\theta$, two triangular end regions of projected length $(25/\tan\theta)\sin\theta$ lie outside the physical filament. Each of these regions contributes only one quarter of the local stacking volume, which leads to the bias fraction given above.}

Filaments with $\tan \theta < 25 / L$ exhibit extreme projection effects in which the projected stacking region significantly exceeds the filament’s physical extent. To mitigate such contamination, we exclude these filaments from our sample, resulting in the removal of 3177 filaments from the full catalog.

After applying both inclination and bias corrections, the relationship between the observed and corrected signal becomes:
\begin{equation}
y_{\mathrm{mod}} = \frac{y_{\mathrm{obs}} \cdot \sin \theta}{1 - \mathrm{bias}}\,,
\label{eq:signal_correction}
\end{equation}
where $y_{\mathrm{mod}}$ is the corrected signal compatible with the idealized perpendicular model. The results in Figure~\ref{fig:bias_correction} demonstrate that our refined geometric bias correction yields a systematically higher signal amplitude compared to inclination-only corrections. The green curve, based solely on $\theta$ correction, tends to underestimate the Compton-$y$ signal, suggesting that the simplified approach used in Tanimura et al.~(2020)~\cite{Tanimura2020A&A...637A..41T} may lead to a systematic underestimation of the thermal contribution from filaments.

We note that this correction is still approximate and subject to simplifying assumptions. In particular, our definition of filament orientation based on endpoints does not capture the full three-dimensional curvature of the filament, which may consist of multiple segments with varying directions. A more accurate approach would involve segment-wise orientation averaging or spline-based modeling of the filament spine. Nonetheless, the correction presented here enhances the consistency between the observations and the model, enabling a more physically meaningful interpretation of the stacked signal.

\bibliographystyle{JHEP}
\bibliography{references_clean3} 
\end{document}